\documentclass[preprint,nofootinbib,aps,superscriptaddress]{revtex4}
\usepackage{color}
\usepackage{mathtools}
\usepackage{array}
\usepackage{booktabs} 
\usepackage{tabularx}
\usepackage{diagbox}
\usepackage{tikz}
\usetikzlibrary{decorations.pathmorphing}
\usepackage{bm}
\usepackage{amsmath,bm,amssymb,amsfonts,dcolumn,color,graphicx,latexsym,epsfig}
\usepackage{rsfso}
\usepackage[pagebackref=false, colorlinks=true]{hyperref}
\definecolor{reddish}{rgb}{0.9,0.3,0.0}  
\definecolor{blueish}{rgb}{0.1,0.1,1}
\hypersetup{
colorlinks=true,
linkcolor=reddish,
filecolor=reddish,      
urlcolor=magenta,
citecolor=blueish,
}
\usepackage{subfigure}
\usepackage{multirow}
\usepackage{natbib}
\usepackage{float}
\usepackage{booktabs}
\usepackage{pifont}
\usepackage{mathrsfs}
\usepackage{caption}
\usepackage{gensymb}
\usepackage{caption, threeparttable}
\usepackage{orcidlink}
\begin{document}
\title{The Penrose process in rotating regular black holes}
\author{Anjan Kar\,\orcidlink{0009-0009-8623-3596}}
\email[Email address: ]{anjankar.phys@gmail.com}
\affiliation{Department of Physics, Indian Institute of Technology, Kharagpur 721 302, India}
\affiliation{Chennai Mathematical Institute, H1, SIPCOT IT Park, Siruseri,
Kelambakkam 603103, India}

%%%%%%%%%%%%%%%%%%%%%%%%%%%%%%%%%%%
\author{Ayan Dey\,\orcidlink{0009-0004-1230-2572}}
\email[Email address: ]{ayandeycob@gmail.com}
\affiliation{Department of Physics, University of Florida, 2001 Museum Road, P.O. Box 118440, Gainesville, FL 32611-8440, USA}

%%%%%%%%%%%%%%%%%%%%%%%%%%%%%%%%%%%
\author{Sayan Kar\,\orcidlink{0000-0001-5854-6223}}
\email[Email address: ]{sayan@phy.iitkgp.ac.in}
\affiliation{Department of Physics, Indian Institute of Technology, Kharagpur 721 302, India}
\begin{abstract}
\noindent We investigate the Penrose process of energy extraction in the context of rotating regular black holes. Previously known geometries, as well as a new rotating regular black hole generated from a recently proposed (Gen. Rel. Grav. {\bf 56}, 52 (2024)) non-rotating, seed solution are considered in our analysis of energy extraction. For the known Neves-Saa class of rotating regular black hole solutions, which includes the Bardeen, Hayward and Fan-Wang spacetimes as special cases, the extraction efficiency is bounded above by the known value for the singular Kerr spacetime. However, in the case of the new rotating regular black hole, which does not have a Schwarzschild singular limit for zero rotation, the extraction efficiency can indeed become much larger, as evident through our results here. Here too, the highest efficiency occurs for the singular solution. Thus, expectedly, both the regularisation parameter in the metric functions, as well as the character of the seed non-rotating solution influence the extent of energy extraction, relative to the respective singular scenarios.
\noindent 
\end{abstract}


\maketitle

\newpage

\section{Introduction}
\noindent A remarkable novelty in rotating black hole spacetimes is the presence of
ergo-regions \cite{Wald} which arise as a consequence of the event horizon and the
infinite redshift surface being different in such spacetimes. The existence of
the ergo-region and its peculiar characteristics with respect to particle trajectories
was noticed and analysed by Penrose and Floyd many years ago \cite{Penrose}. The analysis led to the theoretical proposal of energy (rotational) extraction from a rotating black hole. The basic idea behind this intriguing possibility may be briefly stated in words, as follows. In Kerr (or any rotating black hole spacetime with an ergoregion), particles with negative energy (as seen by an asymptotic observer) can exist in the ergoregion -- a
feature which forms the basis for the Penrose process. In particular, consider the
situation where a particle which enters the ergoregion decays into two particles. Assume one of them has negative energy while for the other, the energy is positive. Therefore, conservation of energy implies that the particle with positive energy, which may escape to infinity, must have more energy than that of the original particle before its decay. This leads to a net energy extraction from the black hole. 
It was further noted that this process could not be eternal since the 
loss of angular momentum of the black hole would eventually halt it.
Technical details on the so-named Penrose process may be found in \cite{Chandra} and 
is also briefly summarised in a later section of this article.

\noindent The Penrose process has been analysed by many authors in the past, in
varying contexts. For instance, one may note variants such as the magnetic Penrose process \cite{ Dhurandhar1, Dhurandhar2, Dhurandhar3, Parthasarathy, Dhurandhar4, Dadhich, Stuchl, Xamidov, DadhichA}, the collisional Penrose process \cite{ZaslavskiiA, BambiA, Hejda, Sang, Bejger}, Penrose process in non-Kerr geometries \cite{Ghosh, Liu, Fatima, Long, Ganguly, Nozawa, Pradhan} and
several other scenarios as discussed in numerous articles \cite{Takahashi, Zaslavskii1, Gupta, Bardeen_Press, Wald1, Blandford, Narayan, ZaslavskiiZ, Das}. 
Further, we also have a host of other, similar processes not just in rotating spacetimes but also in charged, non-rotating black holes \cite{Tursunov, Zaslavskii, Kokubu, Alloqulov, Chen, Vertogradov}.

\noindent In our work here, we analyse the Penrose process in rotating regular black holes
of various types. A regular black hole is considered a potential solution to the spacetime singularity problem \cite{Hawking1, Penrose1, Hawking2} in Einstein's General Relativity (GR).
Regular black holes were first proposed by Bardeen in 1969 as a nonsingular alternative to the singular Schwarzschild solution~\cite{Bardeen}.
Thereafter, a wide range of regular solutions appeared in the literature, including the Hayward solution~\cite{Hayward}, the Dymnikova solution~\cite{Dymnikova1}, and many other models in different theoretical frameworks~\cite{Ayon1, Roman, Dymnikova2, Bronnikov3, Carballo2, Carballo3, Frolov3, Balart1, Bronnikov11, Borissova, Bueno, Eichhorn, Ovalle, Kar2}.
A characteristic property of regular black holes is the presence of one or more horizons while avoiding the central curvature singularity, which is usually replaced by an effective de Sitter or Minkowski core. Despite significant progress, there are several issues in regular black holes which require serious attention. For example, among others, regular black hole thermodynamics
needs to be understood better and is under active investigation~\cite{ma-zhao:2014}.

\noindent Our purpose here is to
investigate how energy extraction efficiency varies
with the so-called regularisation parameter (denoted here by $g$) as well as the
character of the seed non-rotating solution, in different types of
rotating regular black hole spacetimes. 
In recent times, there has been progress in this direction. In particular,  Toshmatov \textit{et al.} \cite{Toshmatov} showed that for the rotating Ay\'on--Beato--Garc\'ia regular metric \cite{Ayon1}, the maximum efficiency is lower than that of the Kerr black hole. 
Patel \textit{et al.} \cite{Patel} reported that for some specific conformally transformed regular black holes, the maximum efficiency can surpass the value for the Kerr case. Although these results suggest that regularising a metric significantly affects energy extraction, it is not clear which properties of a specific rotating regular black hole determines whether the efficiency is enhanced or reduced w.r.t. the corresponding
singular case. 

\noindent In the work reported here, we consider energy extraction in various rotating regular black holes, beginning first with the rotating versions of the Neves--Saa family of solutions \cite{Neves}
and its special cases such as the well-known rotating Bardeen \cite{Bardeen}, Hayward \cite{Hayward} and Fan-Wang \cite{Fan} spacetimes, all of which, in the $g=0$ limit, reduce to Kerr spacetime. Subsequently, we
construct  a new rotating regular geometry which does not possess a Kerr limit
when $g=0$ or the Schwarzschild limit when the rotation is set to zero. Energy
extraction is calculated for this new rotating regular geometry. We find that for the Neves-Saa class, the efficiency is limited above by the Kerr value. However, for the new rotating regular geometry, it exceeds the Kerr case but is limited above by the maximum value for the corresponding singular case.

\noindent Our article is organised as follows.  We first briefly recall the details of
rotating regular black holes in Section II and, thereafter, review the energy extraction process in Section III.
The analysis in Section IV highlights the energy extraction efficiency in a host of rotating regular black hole spacetimes, generically of the Neves-Saa type. 
In Section V, we construct the rotating version of the recently proposed regular black hole which does not have a Schwarzschild limit (for $g=0$) and find the energy extraction efficiency for such rotating geometries. We conclude in Section VI by briefly summarising our findings.
Throughout the paper we adopt the metric signature $(-,+,+,+)$ and set $G = c = 1$.

\section{Rotating regular black holes} 
\noindent In this section, we provide a brief review of static, spherically symmetric regular black holes and discuss the construction of their rotating counterparts as well as their characteristic properties. We primarily focus on a specific class of regular black hole solutions defined by the following spherically symmetric, static line element,
\begin{equation}\label{0.1}
    ds^2=-\left(1-\frac{2m(r)}{r}\right)dt^2+\frac{dr^2}{1-\frac{2m(r)}{r}}+r^2(d\theta^2+\sin^2{\theta}d\phi^2)
\end{equation}
where, the mass function, $m(r)$ is given as
\begin{equation}
    m(r)=\frac{Mr^p}{(r^q+g^q)^{\frac{p}{q}}}
\end{equation}
In the above, $g$ is the regularisation parameter, $M$ denotes the ADM mass and $p$, $q$ are a pair of additional parameters. The various choices of $p$ and $q$ correspond to different black hole geometries. To ensure regularity (i.e., the absence of curvature singularities), it is required that $p\geq 3$ \cite{Fan}. In the limit $g=0$, the above solutions reduce to Schwarzschild geometry.
The well-known Bardeen \cite{Bardeen}, Hayward \cite{Hayward}, and Fan-Wang \cite{Fan} regular black holes are obtained for $p=3$, $q=2$, $p=q=3$ and $p=3$, $q=1$, respectively. This generalisation, representing a wider class of regular black holes was first introduced by Neves and Saa \cite{Neves}, and further discussed in \cite{Fan}.
It is important to mention that the behaviour of the metric functions for this class of spacetimes, for small values of $r$, depends on the value of $p$. The leading order term of the mass function near 
$r\rightarrow 0$ takes the following form
\begin{equation}
    m(r)\to c\, r^p\,,
\end{equation}
 where $c$ is a positive constant. We observe that for $p=3$, the metric function becomes de Sitter ($-g_{tt}=1-cr^2= g^{rr}$) spacetime near the origin, implying that the $r\rightarrow 0$ region is effectively replaced by a de Sitter core. Thus, we may say that the repulsive nature of the de Sitter core prevents the formation of the central singularity. However, for $p>3$, the geometry is no longer de Sitter near the centre ($r\rightarrow 0$), since the metric functions contain terms of powers higher than $r^2$. However, we have checked that all curvature scalars do remain finite, and the spacetime is regular everywhere, and asymptotically flat ~\cite{Fan}. This distinction between the solutions for $p>3$ and $p=3$ can also be understood from the diagonal elements of the required stress-energy tensor, using the Einstein field equations $G_{\mu\nu}=\kappa T_{\mu\nu}$. For $p=3$, the energy density and pressures approach constant values as $r\to 0$, and we have
\begin{equation}
    -T_{t}^{t}=-T_{r}^{r}=\rho_{0},\,\,\,\,\,T_{\theta}^{\theta}=T_{\phi}^{\phi} = -\rho_0
\end{equation}
which is the effective stress energy for de Sitter space, where $\rho_0$ is a constant. However, for $p>3$, all components of the stress energy tensor vanish when $r\to 0$. Thus, the central singularity is replaced by a Minkowski core rather than a de Sitter core when $p>3$. In addition, we note that for
$p>3$, the Weak Energy Condition is violated over a small region near $r\rightarrow 0$, though matter is absent at $r=0$. A good discussion on regular black holes with Minkowski core can be found in Ref.~\cite{new1}.

\noindent From the astrophysical point of view, black holes (or compact objects) are expected to possess spin. Therefore, to test a regular black hole metric with observations, it is essential to construct its rotating counterpart. The rotating version of the above class of stationary solutions was first introduced by Bambi and Modesto \cite{Bambi}. They used the Newman-Janis algorithm \cite{Newman, Newman2} to construct the rotating solution.
Depending on how the `complexification procedure' is incorporated, this method results in two types of rotating solutions. However, in subsequent studies, the type-I solution (as defined by Bambi and Modesto \cite{Bambi}) has been widely adopted as the rotating counterpart. This work is further extended in \cite{Neves} by including the effect of a nonzero cosmological constant $\Lambda$. Useful discussion on rotating regular black holes may be found in \cite{Torres, Torres1}. 

\noindent The type-I rotating regular black hole geometry is described by the following line element,
\begin{equation}\label{0.3}
\begin{split}
    \text{d}s^2 &= -\left(1- \frac{2m(r) r}{\Sigma}\right)\text{d}t^2- \frac{4m(r) r a\sin^2\theta}{\Sigma} \text{d}t\text{d}\phi + \frac{\Sigma}{\Delta} \text{d}r^2 + \Sigma \text{d}\theta^2 \\
    &+ \sin^2\theta \left( r^2 + a^2 + \frac{2m(r) r a^2\sin^2\theta}{\Sigma} \right)\text{d}\phi^2,
    \end{split}
\end{equation}
where $\Sigma=r^2+a^2 \cos^2{\theta}$, $\Delta=r^2-2m(r)r+a^2$ and $m(r)$ is the mass function defined earlier.
The metric reduces to the Kerr metric when the regularisation parameter $g$ vanishes.
Various choices of the parameters $p$ and $q$ lead to different rotating regular black holes. Consequently, we can characterise each rotating regular black hole by the parameter triplet $(p,q,a)$, with $a$ being the rotation parameter. For instance, $(3,2,a)$, $(3,3,a)$ and $(3,1,a)$ correspond to the rotating Bardeen, Hayward and Fan-Wang regular black holes, respectively when $a\neq 0$.
Though the matter required to support the non-rotating versions ($a=0$) of such static, spherically symmetric rotating regular black holes do satisfy the weak energy condition, a finite value of
the rotation ($a\neq 0$) leads to a violation of WEC, as shown in \cite{Bambi, Neves}.

\noindent Rotating regular black holes possess horizons, which obviously are null hypersurfaces. The location of these horizons is determined by the roots of the equation $\Delta=r^2-2m(r)r+a^2=0$ \cite{Poisson}. The number of real positive roots of the equation depends on the spin parameter $a$ and the regularisation parameter $g$. Depending on their values, the spacetime may have two distinct horizons or a single horizon (extremal), or, in some cases, no horizon at all. 
Since we are dealing with rotating spacetimes, rotational motion due to the spin of the central black hole leads to the well-known frame-dragging effect.
This effect causes local inertial frames to be dragged along with the rotation of the black hole. As a result, there exists a region outside the event horizon, bounded by the so-called {\em static limit surface}, where the frame-dragging is so strong that no particle there can remain stationary with respect to the observer at infinity.
This boundary coincides with the event horizon at the poles but extends farther from the black hole at the equator, resulting in an oblate shape.
The region between the event horizon and the {\em static limit surface} is known as the ergo-region.
For various choices of the metric parameters, the extent of the ergo-region changes. 
In the above-mentioned class of rotating regular spacetimes, the {\em static limit surface} remains defined by the usual equation $g_{tt}=0$ \cite{Poisson}.
\begin{figure}[!htbp]
\centering
\subfigure[\hspace{0.1cm}$g=0.4M$ and $a=0.7M$]{\includegraphics[width=0.326\textwidth]{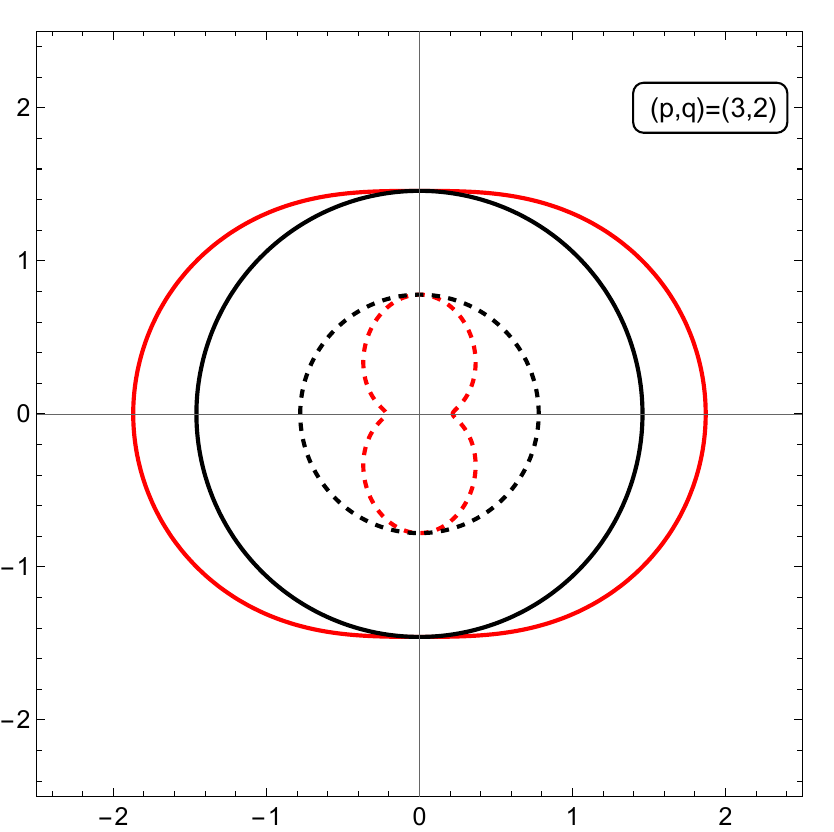}\label{subfig:32-1}}
\subfigure[\hspace{0.1cm}$g=0.4M$ and $a=0.4M$]{\includegraphics[width=0.326\textwidth]{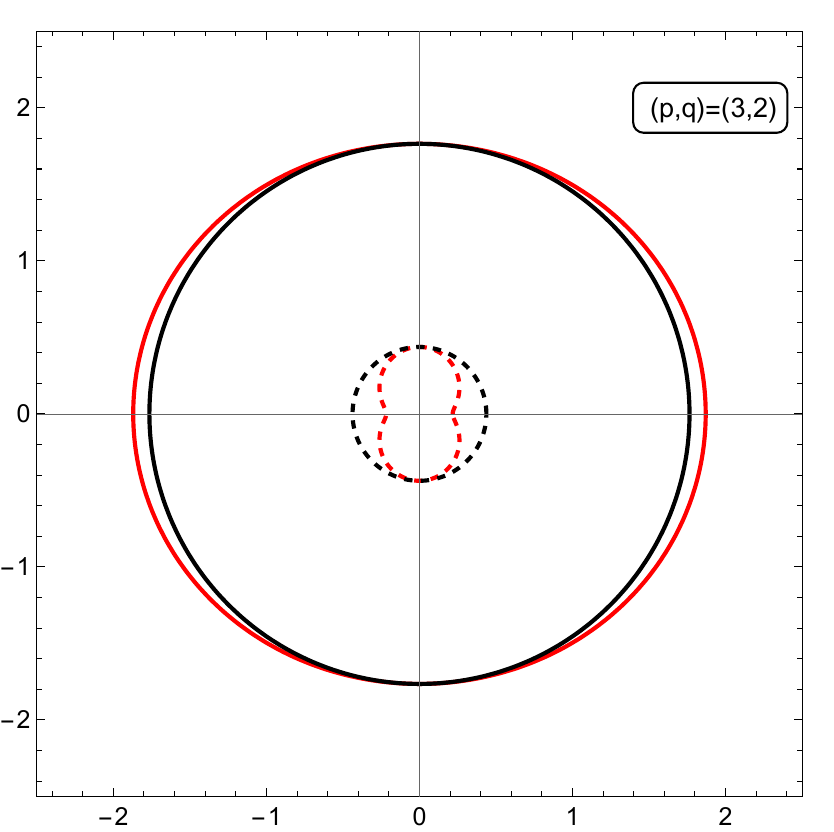}\label{subfig:32-2}}
\subfigure[\hspace{0.1cm}$g=0.4M$ and $a=0.2M$]{\includegraphics[width=0.326\textwidth]{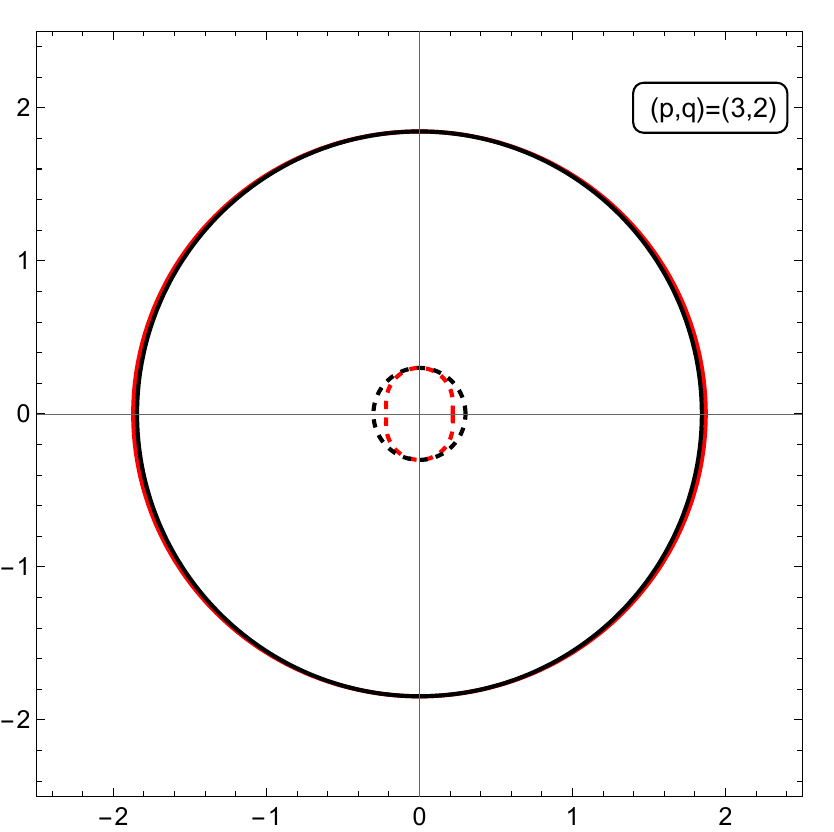}\label{subfig:32-3}}
\subfigure[\hspace{0.1cm}$g=0.4M$ and $a=0.7M$]{\includegraphics[width=0.326\textwidth]{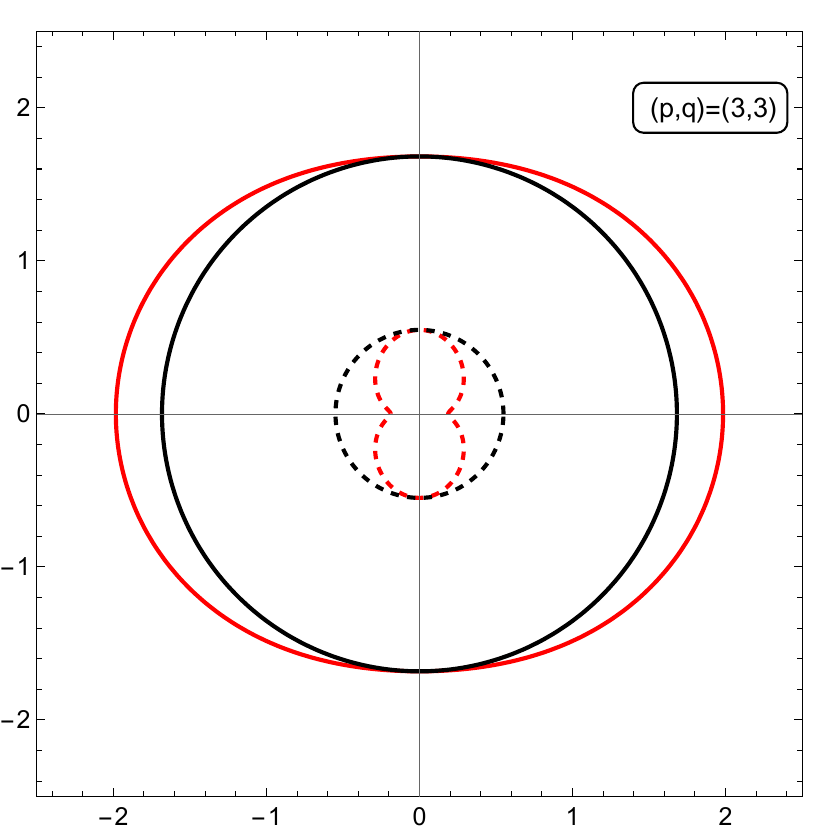}\label{subfig:33-1}}
\subfigure[\hspace{0.1cm}$g=0.4M$ and $a=0.4M$]{\includegraphics[width=0.326\textwidth]{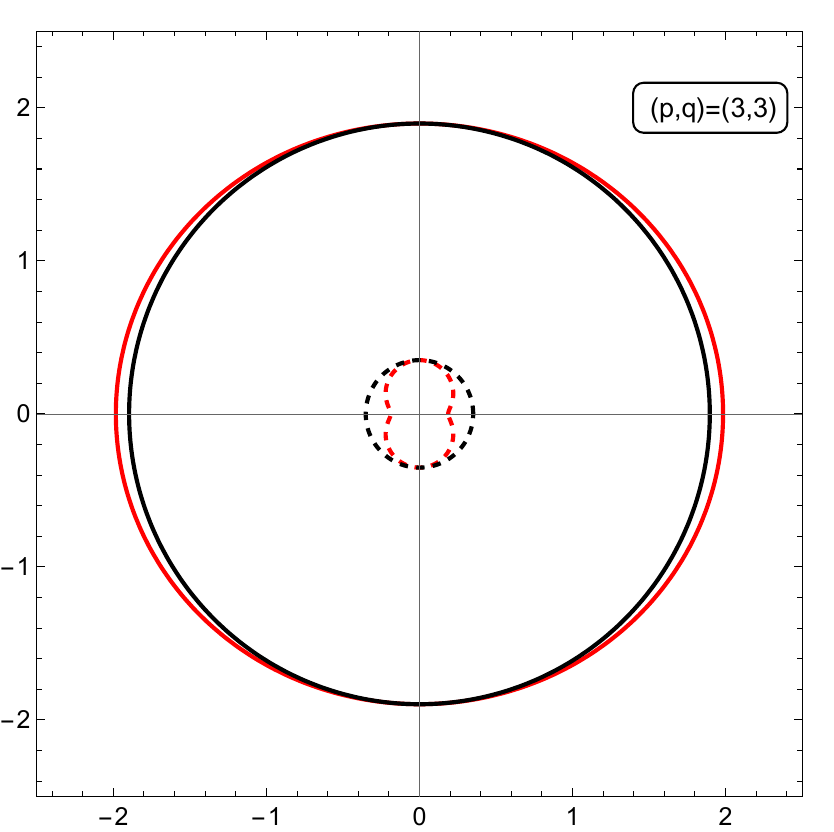}\label{subfig:33-2}}
\subfigure[\hspace{0.1cm}$g=0.4M$ and $a=0.2M$]{\includegraphics[width=0.326\textwidth]{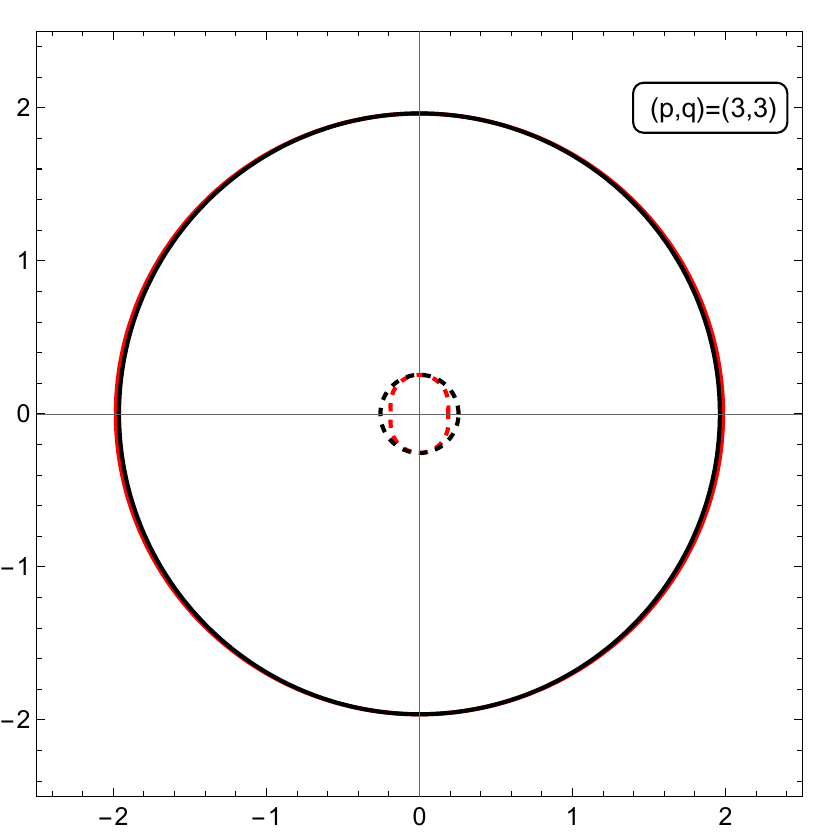}\label{subfig:33-3}}
\subfigure[\hspace{0.1cm}$g=0.1M$ and $a=0.7M$]{\includegraphics[width=0.326\textwidth]{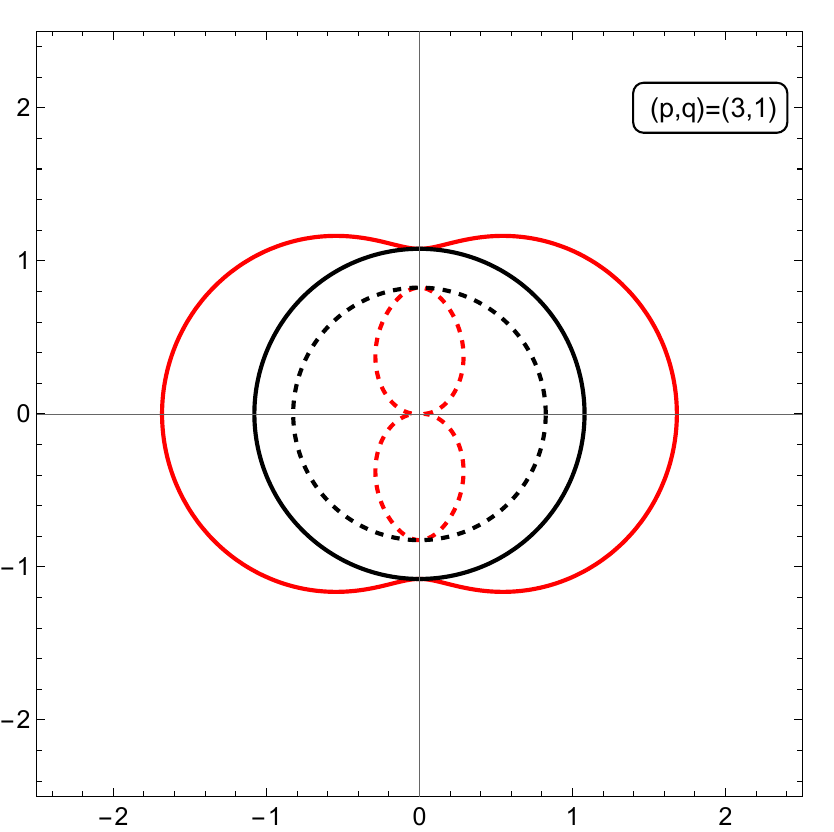}\label{subfig:31-1}}
\subfigure[\hspace{0.1cm}$g=0.1M$ and $a=0.7M$]{\includegraphics[width=0.326\textwidth]{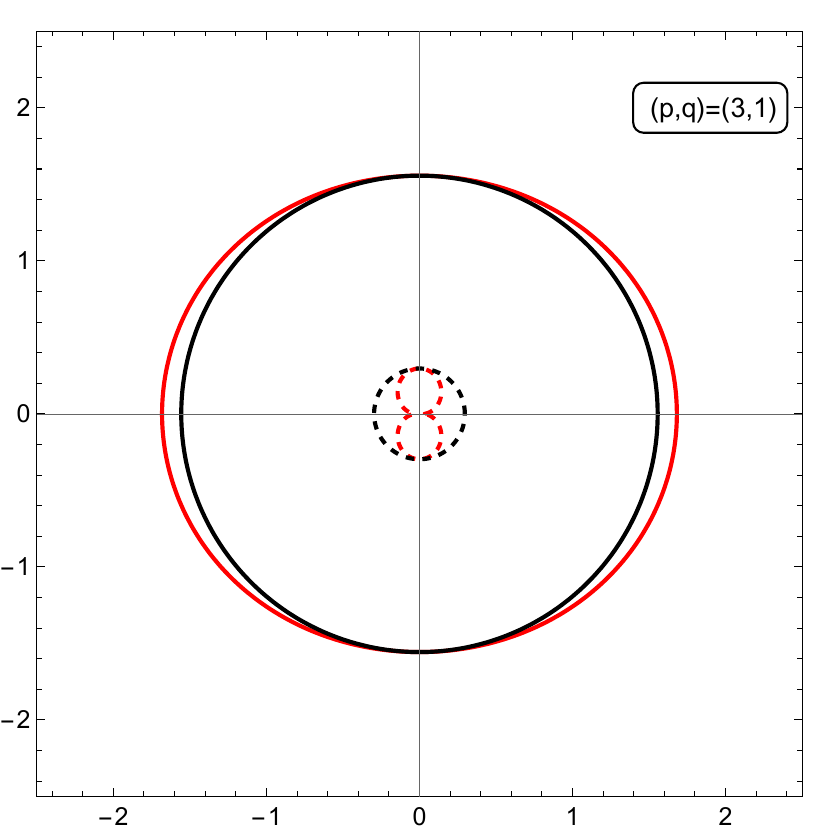}\label{subfig:31-2}}
\subfigure[\hspace{0.1cm}$g=0.1M$ and $a=0.7M$]{\includegraphics[width=0.326\textwidth]{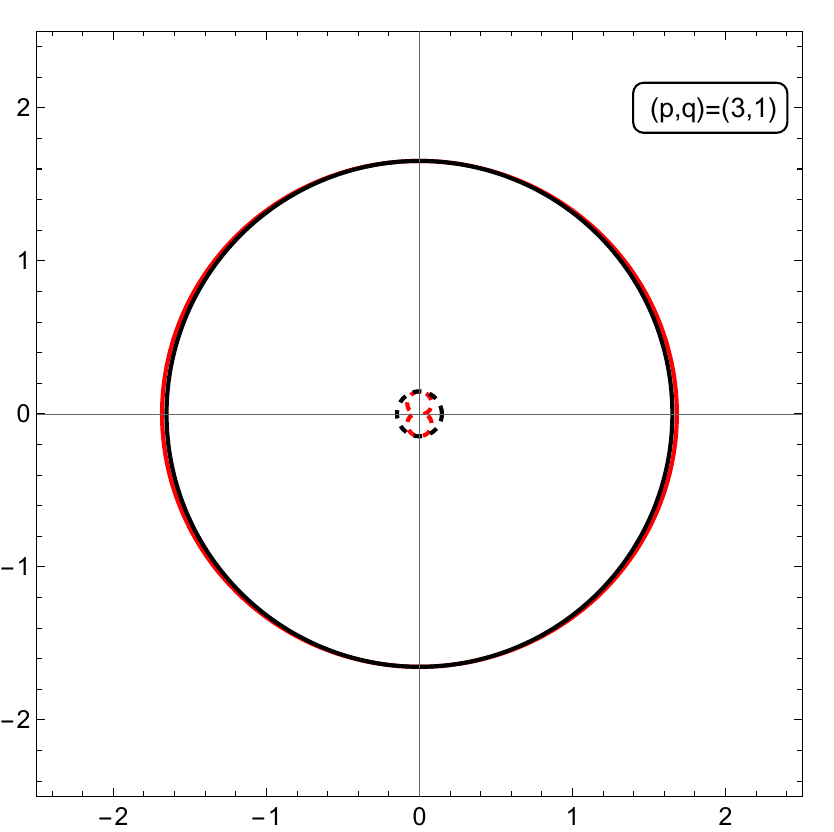}\label{subfig:31-3}}
\caption{Variation of the {\em static limit surface} (solid red) together with the inner and outer horizon (black) of the Bardeen $(3,2,a)$, Hayward $(3,3,a)$ and Fan-Wang $(3,1,a)$ rotating spacetimes for various choices of $a$ and $g$. Solid: outer, Dashed: inner.}
\label{fig:(32,33)}
\end{figure}

\begin{figure}[h]
\centering
\subfigure[\hspace{0.1cm}(p,q)=(3,1)]{\includegraphics[width=0.326\textwidth]{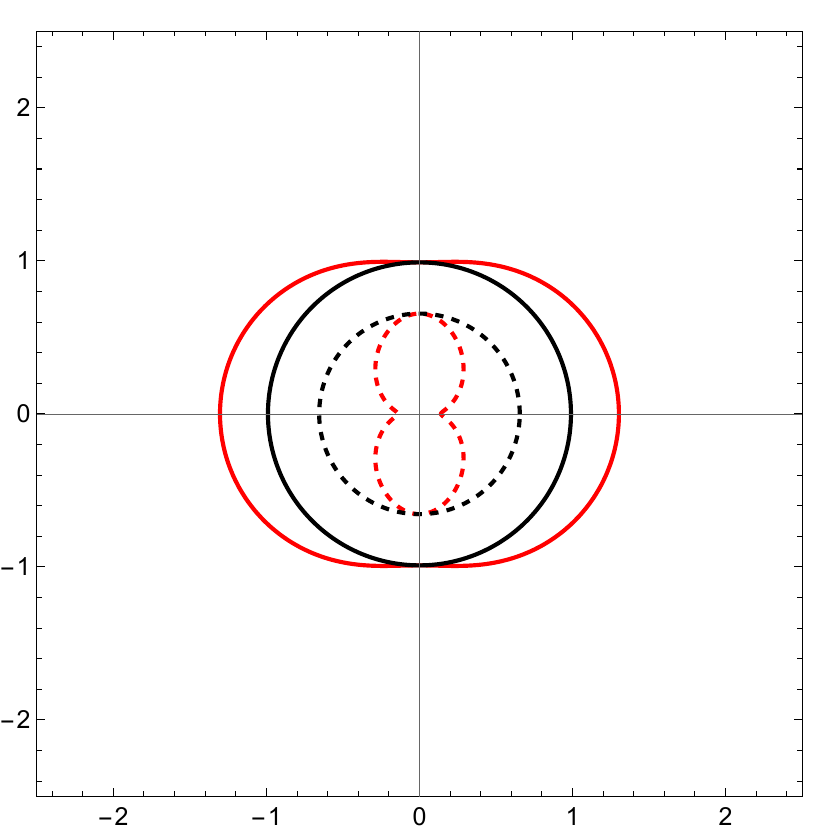}\label{subfig:31}}
\subfigure[\hspace{0.1cm}(p,q)=(3,2)]{\includegraphics[width=0.326\textwidth]{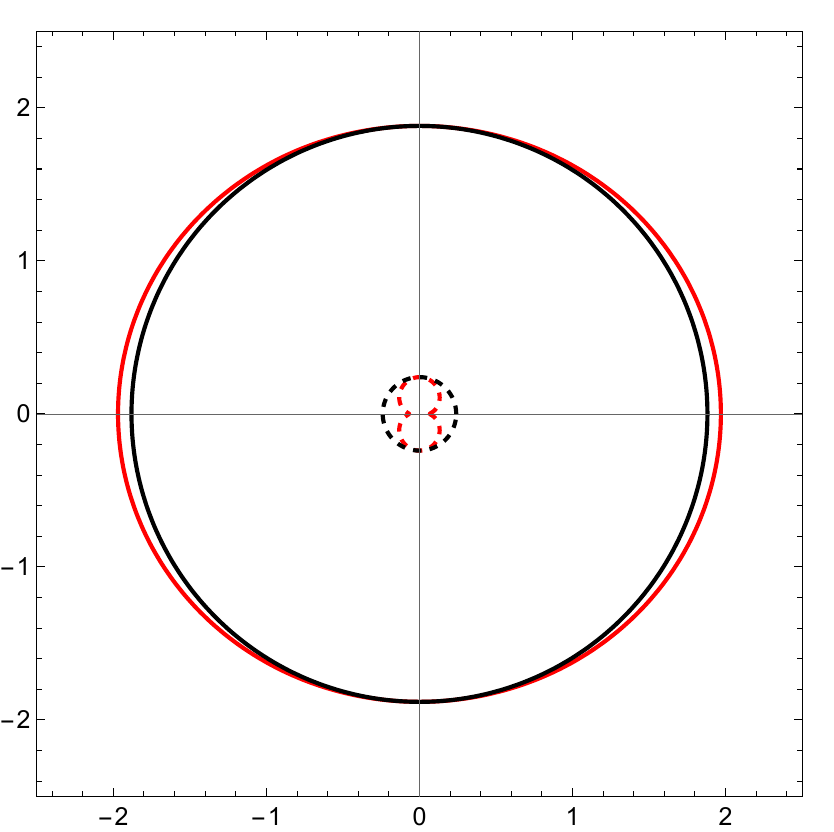}\label{subfig:32}}
\subfigure[\hspace{0.1cm}(p,q)=(3,3)]{\includegraphics[width=0.326\textwidth]{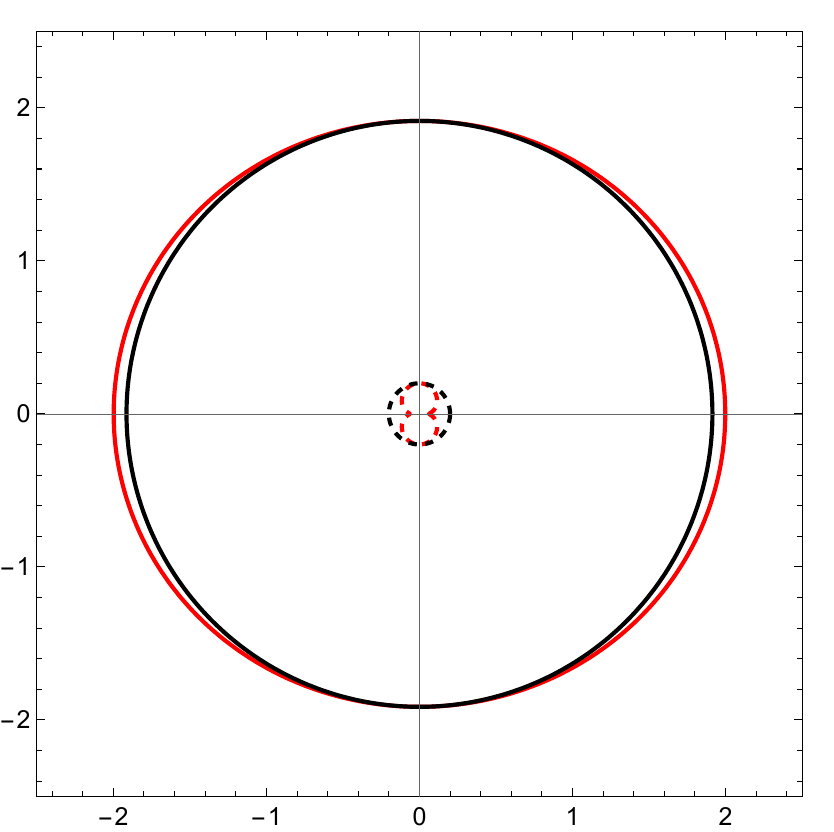}\label{subfig:33}}
\caption{ Variation of ergo-region (region between solid red and solid black) for rotating regular black holes with $p=3$ and various $q$ values is shown, for $a=0.4M$ and $g=0.2M$.}
\label{fig:(3n)}
\end{figure}

\begin{figure}[!htbp]
\centering
\subfigure[\hspace{0.1cm}(p,q)=(3,3)]{\includegraphics[width=0.326\textwidth]{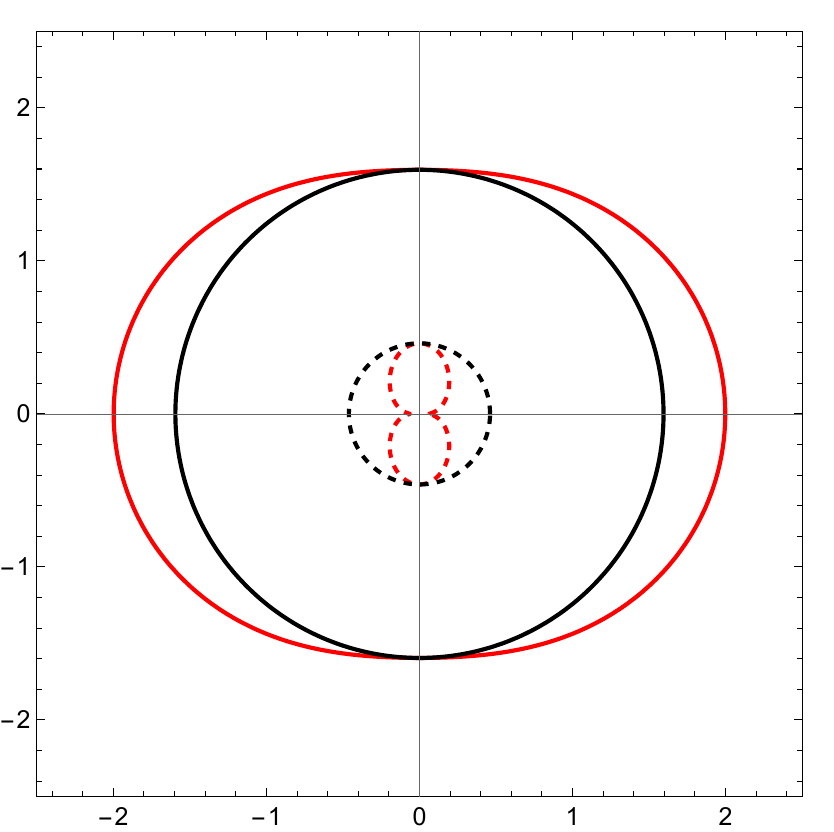}\label{subfig:33--1}}
\subfigure[\hspace{0.1cm}(p,q)=(4,3)]{\includegraphics[width=0.326\textwidth]{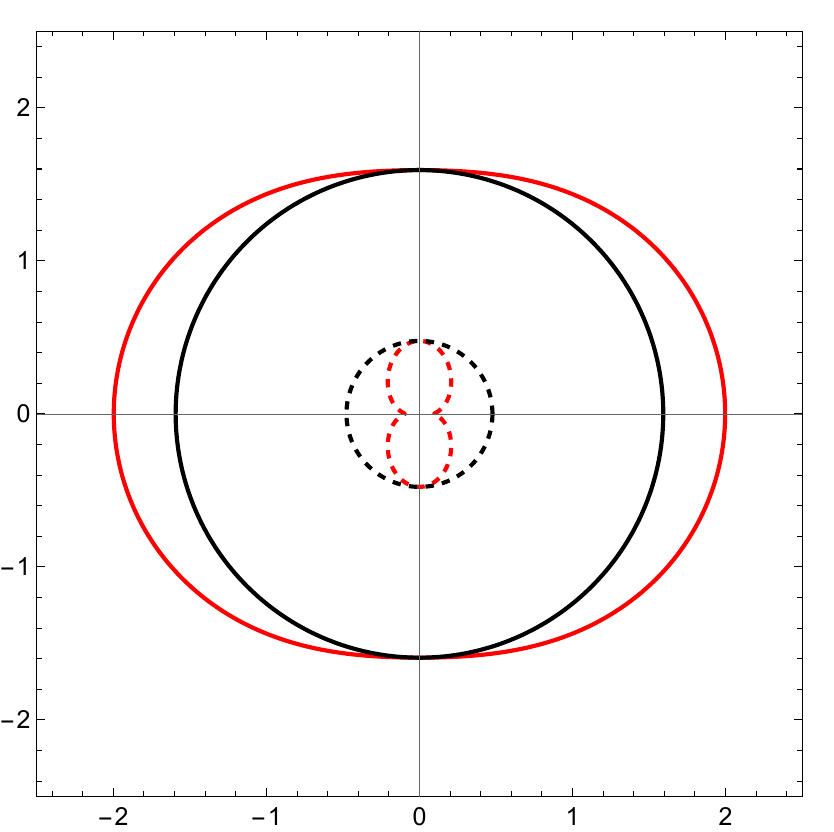}\label{subfig:43}}
\subfigure[\hspace{0.1cm}(p,q)=(5,3)]{\includegraphics[width=0.326\textwidth]{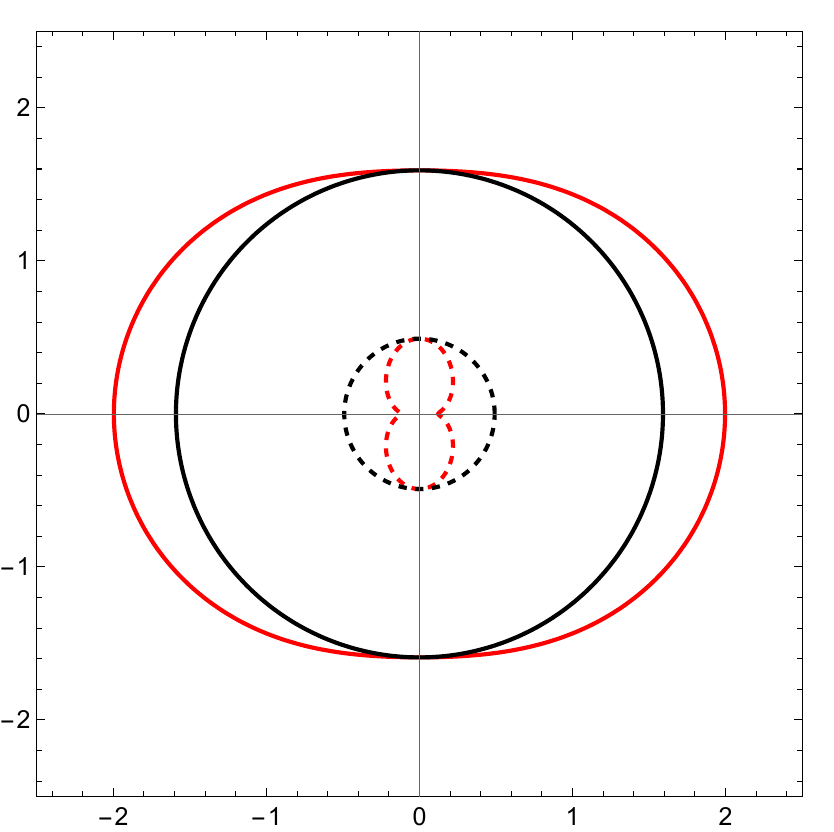}\label{subfig:53}}
\caption{ Graph of ergo-region (region between solid red and solid black) for rotating regular black holes with $q=3$ and various $p$ values, for $a=0.8M$ and $g=0.2M$.}
\label{fig:(n3)}
\end{figure}

\noindent In Figure \ref{fig:(32,33)}, we have shown the different shapes of ergo-regions and horizon radii, for Bardeen $(3,2,a)$, Hayward $(3,3,a)$ and Fan-Wang $(3,1,a)$ rotating metrics corresponding to different combinations of $a$ and $g$. It is clear that for a smaller rotation parameter, the ergo-region is relatively reduced in size, whereas higher rotation increases its width. In Figure \ref{fig:(3n)}, the variation of ergo-region with respect to the parameter $q$ is represented for fixed values of $p$, $a$ and $g$. Similarly,  variations with respect to $p$ for some fixed values of $q$, $a$ and $g$ are illustrated in Figure \ref{fig:(n3)}. It is observed that for the selected values of $a$, $g$, and $p$, the ergo-region shows a significant variation with $q$. In contrast, for fixed $a$, $g$, and $q$, the variation with respect to $p$ is comparatively less pronounced.
In the next section, we will briefly recall rotational energy extraction from rotating black holes via the Penrose process.

\section{Energy extraction from a rotating black hole via the Penrose process}
\noindent Let us now briefly review the Penrose process, a well-known mechanism for extracting rotational energy from black holes \cite{Penrose, Chandra}. The process relies on the existence of an ergo-region, where the frame-dragging effect is sufficiently strong to prevent the possibility of stationary observers, as is well-known and 
also noted earlier here. Let us consider an incident massive particle with energy $E_{(0)}$ and angular momentum $L_{(0)}$ which splits into two massless particles $(E_{(1)}, L_{(1)})$ and $(E_{(2)}, L_{(2)})$ in the ergo-region. One of them, $(1)$, crosses the event horizon with $(E_{(1)}<0)$ and the other, $(2)$, escapes to infinity with energy higher than that of the original test particle (i.e. $E_{(0)}$). Taking into consideration the fact that the particles follow the stationary axisymmetric conservation laws, we have
\begin{equation}\label{0.4}
    E_{(0)}=E_{(1)}+E_{(2)}, \hspace{1cm} L_{(0)}=L_{(1)}+L_{(2)}
\end{equation}
Here, energy and angular momentum are two conserved quantities for a geodesic $u^{\mu}$ in a stationary axisymmetric spacetime, defined as: $E=-g_{\mu\nu}(\partial_t)^{\mu}u^{\nu}$ and $L=g_{\mu\nu}(\partial_{\phi})^{\mu}u^{\nu}$.
To extract the rotational energy via the Penrose process, the condition $E_{(1)}<0$ should be fulfilled. To evaluate how efficient the energy extraction is, we analyse the geodesics of the metric in Eq.(\ref{0.3}) using the Hamilton--Jacobi method \cite{Chandra, Carter, Bhattacharya}, where we have
\begin{equation}\label{0.5}
    2\frac{\partial S}{\partial \lambda}=g^{\mu\nu}\frac{\partial S}{\partial x^\mu}\frac{\partial S}{\partial x^\nu}
\end{equation}
with $\lambda$ being the affine parameter along the geodesics, and $S$ can be written as,
\begin{equation}
    S=-\frac{1}{2}\epsilon \lambda-Et+L\phi+S_{r}(r)+S_{\theta}(\theta)
\end{equation}
Here, $\epsilon=0$ for massless particles and $\epsilon=1$ for massive particles. Upon solving Eq.(\ref{0.5}) with the above ansatz, the left-hand side becomes purely radial and the right-hand side purely angular $(\theta)$, allowing both to be set equal to a separation constant, the Carter constant $(-\zeta)$ \cite{Carter}. 
Moreover, the momenta in the radial and polar directions are given by $p_r=dS_r/dr$ and $p_\theta=dS_\theta/d\theta$ 
respectively, leading to the following radial and polar equations:
\begin{equation}\label{0.7}
    \begin{aligned}
        \Sigma^2\dot{r}^2=(r^2+a^2)^2\left(E-\frac{aL}{r^2+a^2}\right)^2-\Delta(\zeta+r^2\epsilon+(L-aE)^2)\\
        \Sigma^2\dot{\theta}^2=-\frac{1}{\sin^2\theta}(Ea\sin^2\theta-L)^2+(\zeta-\epsilon a^2\cos^2\theta+(L-aE)^2)
    \end{aligned}
\end{equation}
Let us consider a simple scenario where test particles are restricted to the equatorial plane $\theta=\pi/2$ (i.e. $ \zeta=0$). As mentioned earlier, in the Penrose process, the incident massive particle after entering the ergo-region encounters a turning point $(\dot{r}=0)$ and splits into two massless particles. Thus from Eq.(\ref{0.7}) we have, 
\begin{equation}\label{0.8}
    E=\frac{2aLm/r\pm\left\{4a^2L^2m^2/r^2+(r^2+a^2(1+2m/r)(L^2(1-2m/r)+\Delta\epsilon))\right\}^{1/2}}{r^2+a^2(1+2m/r)}
\end{equation}
Or, the angular momentum can be expressed as
\begin{equation}\label{0.9}
    L=\frac{-2aEm/r\pm E\sqrt{\Delta}(1-\epsilon(1-2m/r)/E^2)^{1/2}}{1-2m/r}
\end{equation}
From Eq.(\ref{0.8}), it follows that the positive sign must be chosen to ensure positive energy in the limit $a\to 0$. 
Therefore, a massless particle possesses negative energy when its angular momentum is negative (i.e., counter-rotating with respect to the black hole’s rotation) and the condition $(1-2m/r)<0$ is satisfied. The condition $(1-2m/r)<0$ represents the region inside the ergo-region (as, $g_{tt}(\theta=\pi/2)=1-2m/r$).

\noindent Next, we evaluate the angular momentum of the incident massive particle and the resulting two massless particles using Eq.(\ref{0.9}). We have
\begin{equation}
\begin{aligned}
     L_{(0)}=E_{(0)}\frac{-2am/r+ \sqrt{\Delta}(1-(1-2m/r)/E^{2}_{(0)})^{1/2}}{1-2m/r}\\
      L_{(1)}=E_{(1)}\frac{-2am/r- \sqrt{\Delta}}{1-2m/r}\\
       L_{(2)}=E_{(2)}\frac{-2am/r+ \sqrt{\Delta}}{1-2m/r}
\end{aligned}
\end{equation}
Using the conservation equations in (\ref{0.4}), we have 
\begin{equation}
    \begin{aligned}
        E_{(1)}=-\frac{1}{2}\left\{\left(1-\frac{1-2m/r}{E_{(0)}^2}\right)^{1/2}-1\right\}E_{(0)}\\
        E_{(2)}=\frac{1}{2}\left\{\left(1-\frac{1-2m/r}{E_{(0)}^2}\right)^{1/2}+1\right\}E_{(0)}
    \end{aligned}
\end{equation}
From the above equations, it is clear that energy extraction is only possible (i.e. $E_{(2)}>E_{(0)}$) inside the ergo-region. The extracted energy $(\delta E)$ can be expressed as,
\begin{equation}\label{12}
    \delta E=-E_{(1)}=\frac{1}{2}\left\{\left(1+\frac{a^2-\Delta}{r^2E_{(0)}^2}\right)^{1/2}-1\right\}E_{(0)}
\end{equation}
Therefore, the maximum energy extraction occurs at $\Delta=0$, which represents the horizons of the rotating spacetime.
As a result, the maximum rotational energy may be extracted close to the horizon, which can be expressed as
\begin{equation}\label{0.13}
     \delta E_{\text{max}}=\frac{1}{2}\left(\sqrt{1+\frac{a^2}{r_{h}^2E_{(0)}^2}}-1\right)E_{(0)}
\end{equation}
where $r_h$ represents the horizon radius. For the value $E_{(0)}=1$, the maximum extracted energy for the Kerr black hole is reported as $0.207$ \cite{Chandra}.
Further, we need to note that the above analysis
is valid when $m$ is replaced by an appropriate function $m(r)$-- a fact we use
in the following sections of this article.

\section{Energy extraction efficiency of rotating regular black holes (Neves-Saa class)}
\noindent Having recalled the Penrose process for generic rotating black holes, we now evaluate the maximum energy extraction efficiency for the class of rotating regular black holes discussed in Section II. We have noted in Eq.(\ref{0.13}) the maximum rotational energy which may be extracted in the vicinity of the horizon. Although the rotating regular black holes may have multiple horizon radii, we consider only the outer horizon radius as the relevant location for maximum energy extraction. The inner horizon is causally disconnected from our universe. 
Thus, the generic scenario 
is the following: a massive particle enters the ergo-region from the region
exterior to the outer horizon and splits into two massless particles. One massless particle resides inside the outer horizon with negative energy, while the other emerges with higher energy than that of the original particle before it had split. We now obtain and discuss the extraction efficiency for different regular black holes using (\ref{12}) and (\ref{0.13}). 

\subsection{Rotating Bardeen regular black hole} 
\begin{figure}[!htbp]
\centering
\subfigure[\hspace{0.1cm}Outer horizon]{\includegraphics[width=0.43\textwidth]{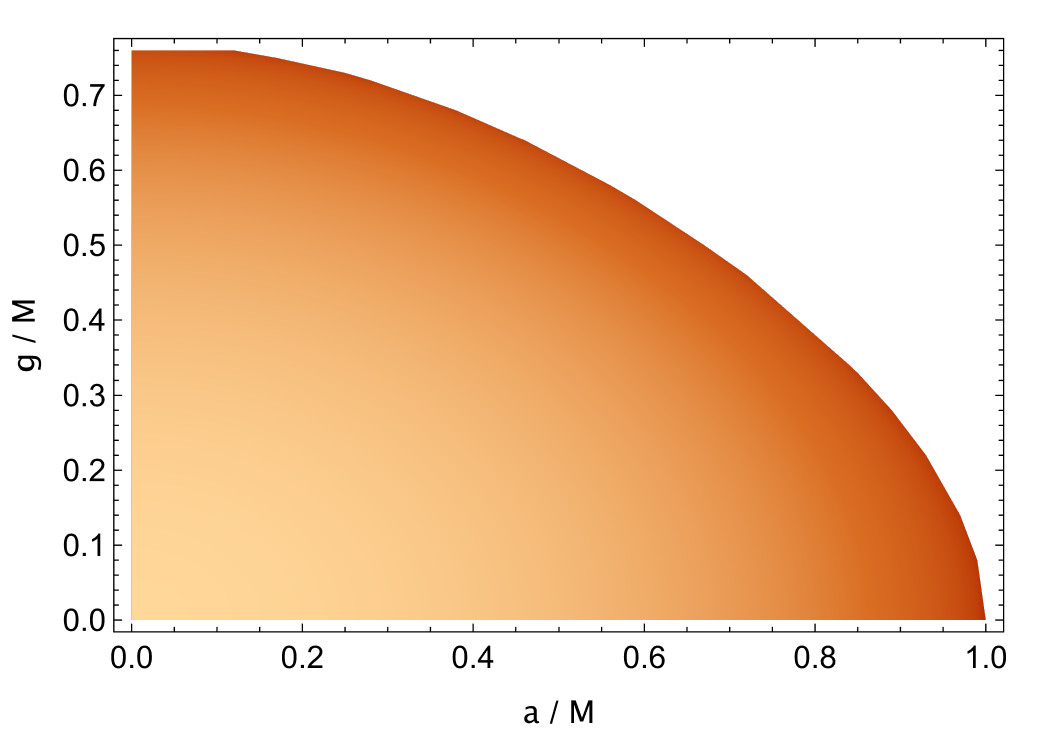}\label{subfig:B_outer}}
\subfigure[\hspace{0.1cm}Maximum efficiency]{\includegraphics[width=0.5\textwidth]{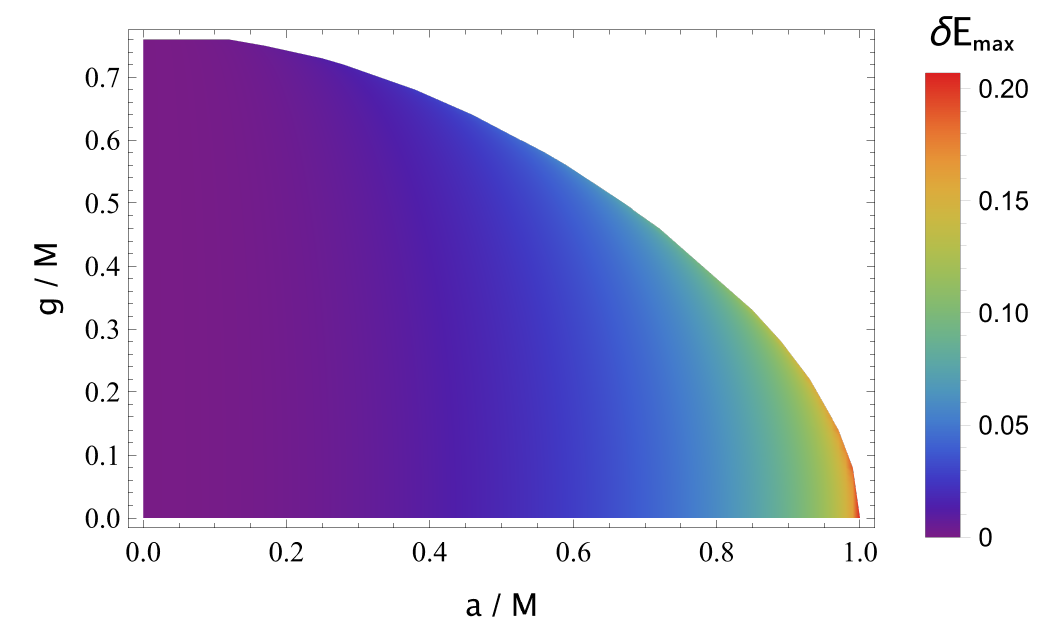}\label{subfig:B_extraction}}
\caption{The allowed values of $a/M$ and $g/M$ for which an outer horizon exists in the rotating Bardeen black hole (left). The right subfigure shows the variation of the maximum energy extraction efficiency with $a/M$ and $g/M$.}
\label{fig:Bardeen}
\end{figure}

\noindent Let us consider the rotating Bardeen metric $(3,2,a)$ which can be identified with the following mass function,
\begin{equation}
    m(r)=M\left(\frac{r^2}{r^2+g^2}\right)^\frac{3}{2}
\end{equation}
In Figure \ref{subfig:B_outer}, we demonstrate the parameter space of $a/M$ and $g/M$ for which an outer horizon exists for the rotating Bardeen black hole. The coloured region of the figure represents the allowed $a/M$ and $g/M$ for which an
outer horizon is present.
The maximum energy extraction efficiency of the rotating Bardeen regular black hole is shown in Figure \ref{subfig:B_extraction}. It is observed that the efficiency is reduced in the presence of a finite $g$ compared to the case $g=0$, for which it is $0.207$.

\subsection{Rotating Fan-Wang regular black hole}
\begin{figure}[!htbp]
\centering
\subfigure[\hspace{0.1cm}Outer horizon]{\includegraphics[width=0.43\textwidth]{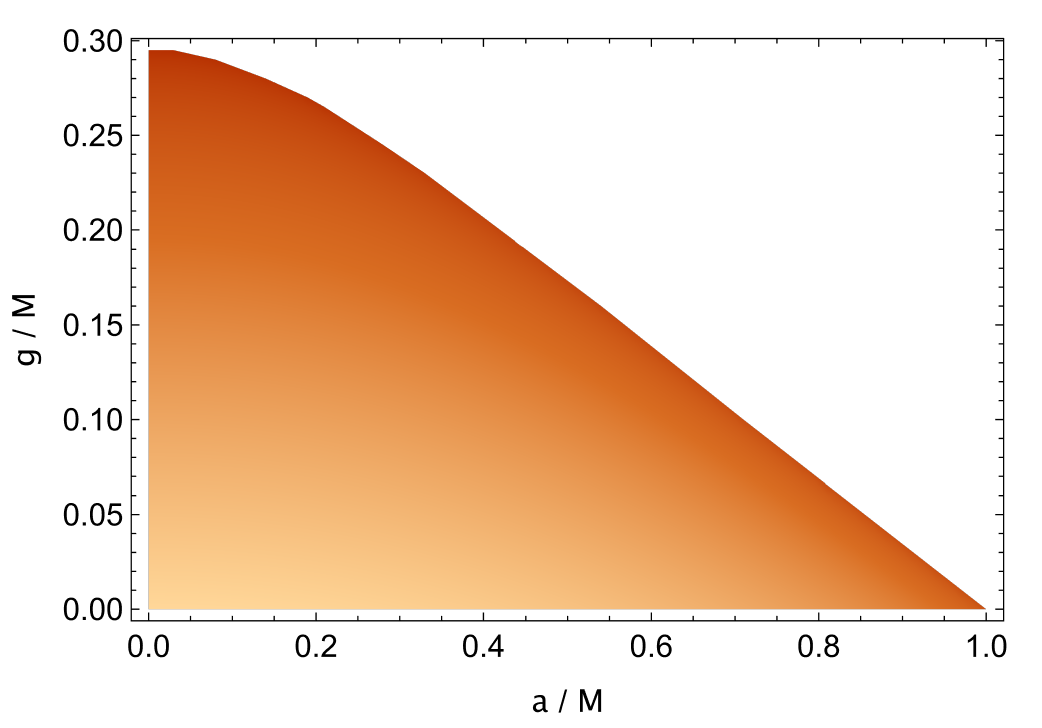}\label{subfig:H_outer}}
\subfigure[\hspace{0.1cm}Maximum efficiency]{\includegraphics[width=0.5\textwidth]{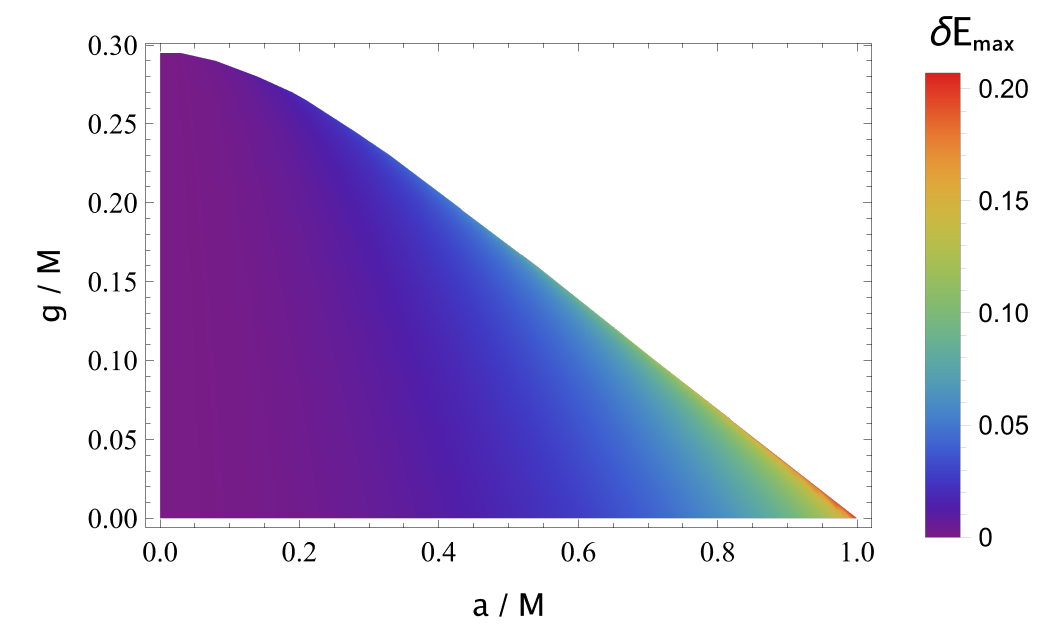}\label{subfig:H_extraction}}
\caption{The allowed values of $a/M$ and $g/M$ for which an outer horizon exists in the rotating Fan-Wang black hole (left). The right subfigure shows the variation of the maximum energy extraction efficiency with $a/M$ and $g/M$.}
\label{fig:Hayward}
\end{figure}

\noindent The rotating Fan-Wang regular black hole $(3,1,a)$ has the following mass function,
\begin{equation}
    m(r)=M\frac{r^3}{(r+g)^3}
\end{equation}
One may note that the region of the parameter space of $a/M$ and $g/M$ having an outer horizon for the above mass function (as demonstrated in Figure \ref{subfig:H_outer}) has an area smaller than that for the rotating Bardeen metric (Figure \ref{subfig:B_outer}). 
The dependency of the maximum energy extraction efficiency $(\delta E_{\text{max}})$ for the rotating Fan-Wang metric is shown in Figure \ref{subfig:H_extraction}. Similar to the rotating Bardeen metric, efficiency reaches the maximum value when $g$ vanishes.

\subsection{Other rotating solutions from the Neves--Saa generalisation}
\begin{figure}[h]
    \centering
    \includegraphics[width=0.492\textwidth]{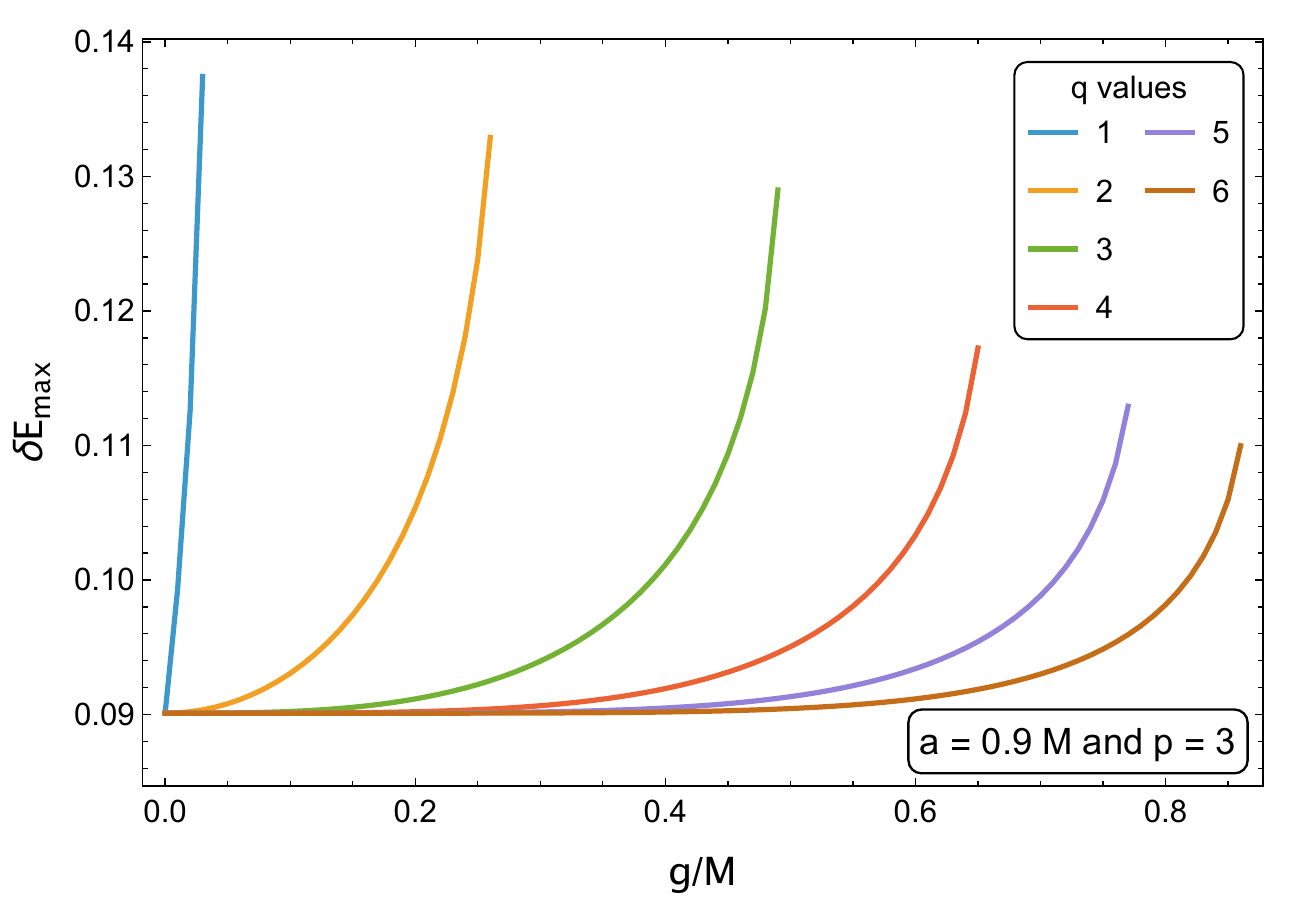}
         \includegraphics[width=0.492\textwidth]{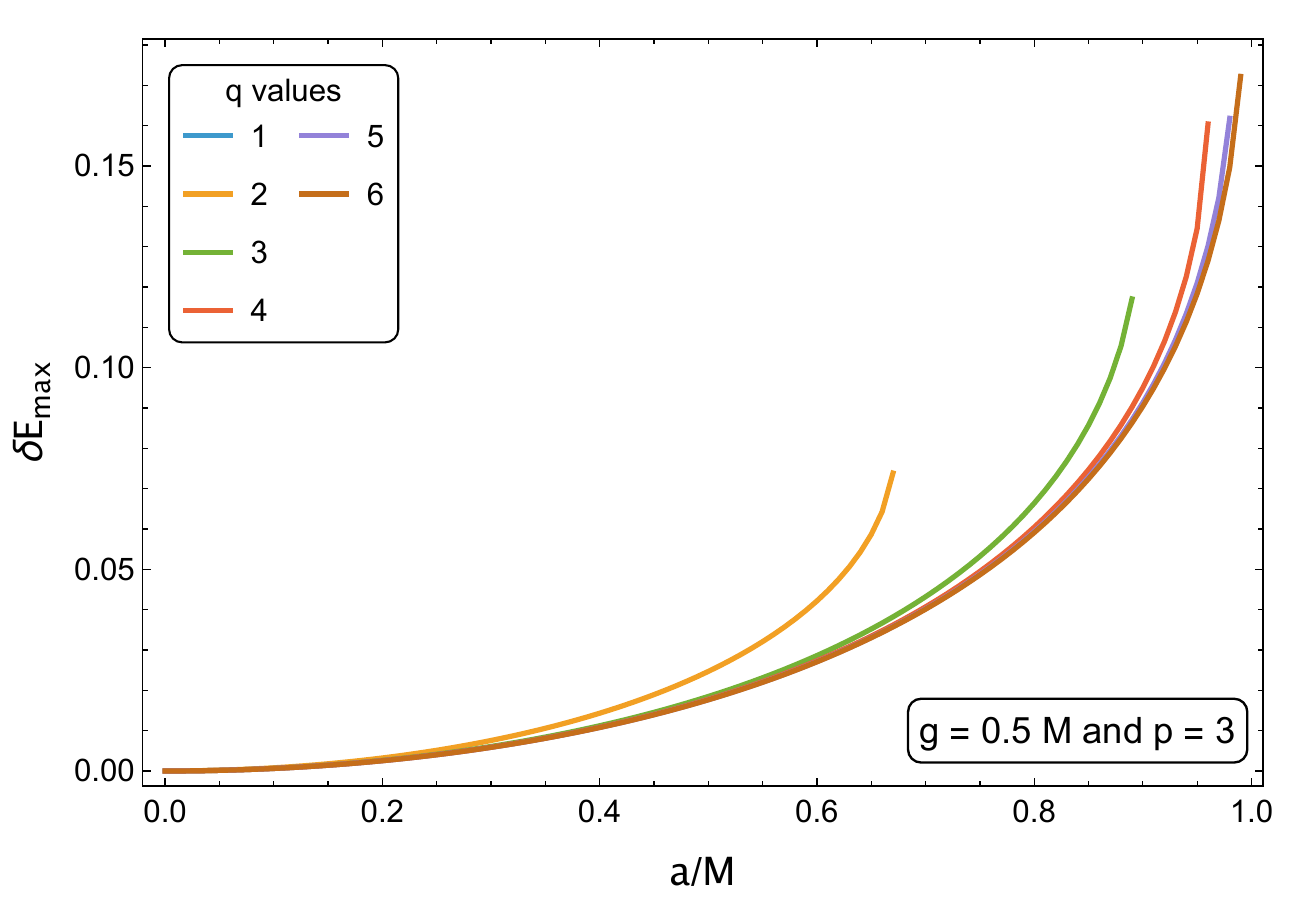}
    \caption{The variation of maximum energy extraction efficiency with $g/M$ (left) and $a/M$ (right) for different values of $q$.}
    \label{fig:NS_variation_q}
\end{figure}

\begin{figure}[h]
    \centering
    \includegraphics[width=0.492\textwidth]{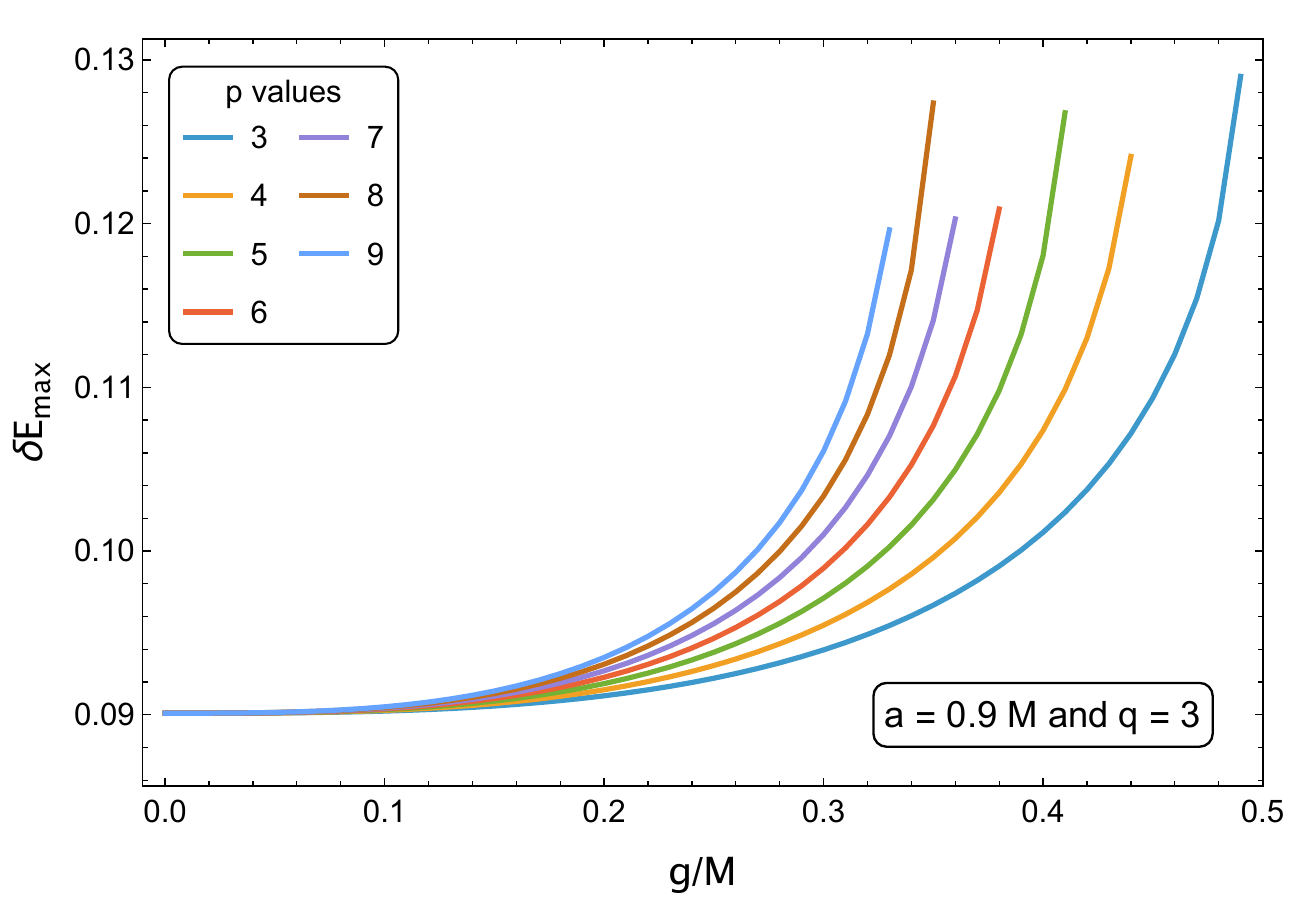}
         \includegraphics[width=0.492\textwidth]{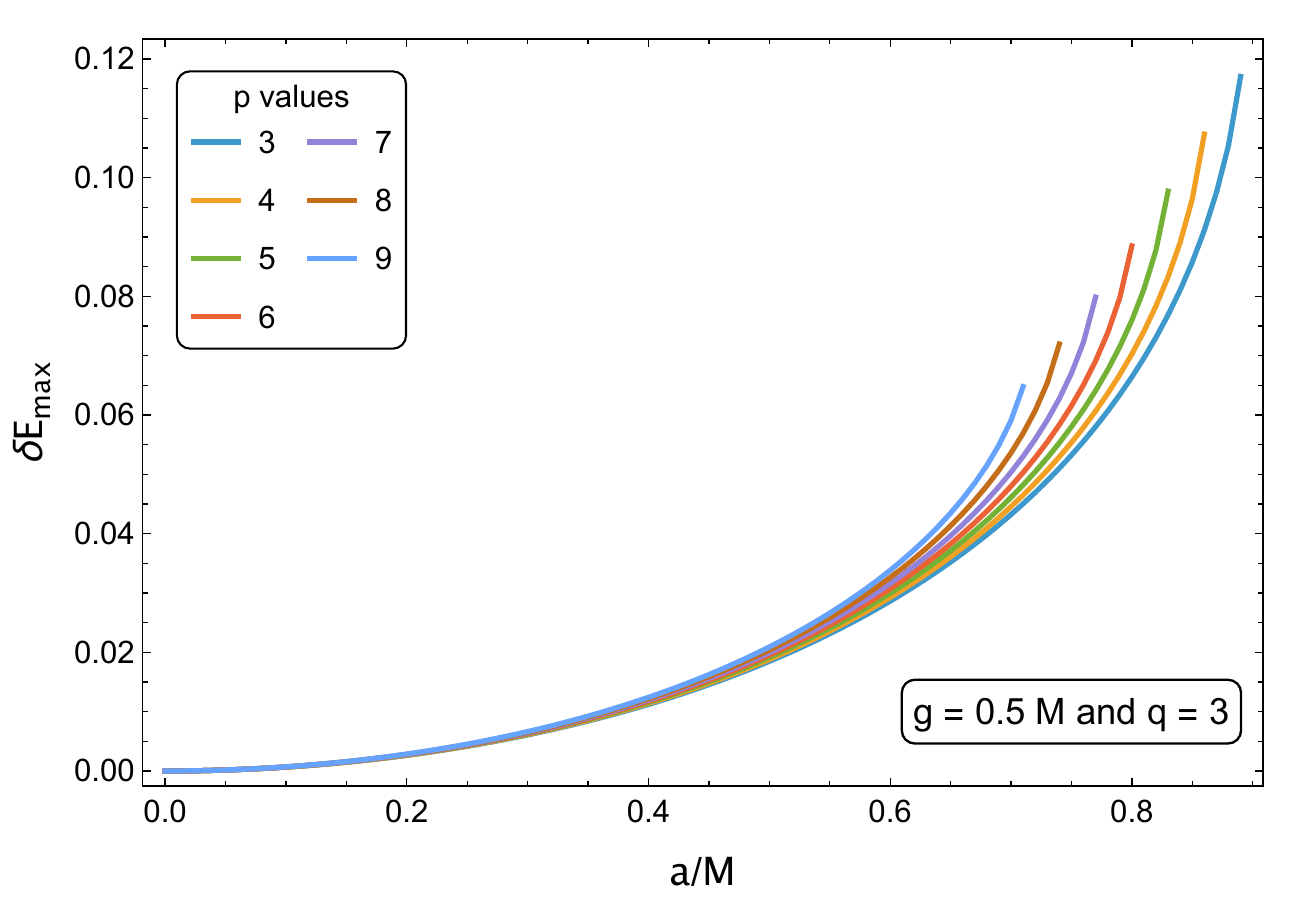}
    \caption{The variation of maximum energy extraction efficiency with $g/M$ (left) and $a/M$ (right) for different values of $p$.}
    \label{fig:NS_variation_p}
\end{figure}

\noindent As mentioned earlier, the generalised Neves and Saa metric $(p,q,a)$ can be identified with the following mass function,
\begin{equation}
    m(r)=M\frac{r^p}{(r^q+g^q)^{p/q}}
\end{equation}
We discussed above the maximum efficiency $(\delta E_{\text{max}})$ for two special cases of the above generalisation, namely Bardeen $(3,2,a)$ and Fan-Wang $(3,1,a)$. 
The maximum efficiency for the rotating Hayward black hole $(3,3,a)$ is similarly found to be constrained by the Kerr limit, in agreement with the results reported in \cite{Khan}. As this case has ulae for likelihood-based tests of new physicsulae for likelihood-based tests of new physicsalready been studied in \cite{Khan}, we do not elaborate on it further here.
We now study how the efficiency depends on other values of the parameters $p$ and $q$.
In Figure \ref{fig:NS_variation_q} (left), the variation of $\delta E_{\text{max}}$ with $g/M$ for different values of $q$ is shown ranging from $q=1$ to $q=6$, with $p=3$ and $a=0.9M$. The corresponding variation with $a/M$ is illustrated in Figure \ref{fig:NS_variation_q} (right), where $p=3$ and $g=0.5M$ are fixed.
One may note that for fixed values of $a$ and $p$, an increase in $q$ 
allows a wider range of $g$ values before the maximum is reached (see Figure \ref{fig:NS_variation_q} (left)). However, the maximum efficiency value is found to decrease with increasing $q$. On the other hand, for fixed values of $g$ and $p$, the efficiency increases with $a$, and this growth is more pronounced for higher $q$ (Figure \ref{fig:NS_variation_q}(right)).
Figure \ref{fig:NS_variation_p} shows the variation of $\delta E_{\text{max}}$ with $g/M$ (for fixed $a/M$ and $q$) and $a/M$ (for fixed $g/M$ and $q$) for different values of $p$ ranging from $p=3$ to $p=9$. We find that for fixed $a$ and $q$, maximum efficiency increases with $g$ (Figure \ref{fig:NS_variation_p} (left)). Similarly, it increases with $a$ for fixed $g$ and $q$ (Figure \ref{fig:NS_variation_p}(right)). In both cases, the rate of increase is more significant for smaller values of $p$ ($p\geq3$).

\noindent Thereafter, in Figure \ref{fig:NS_variation_q}, we study the maximum efficiency of the Penrose process by varying either $a/M$ or $g/M$ separately for different values of $q$, while keeping the other parameters constant. A similar analysis is performed for different values of $p$ in Figure \ref{fig:NS_variation_p}. In each case, the efficiency $\delta E_{\text{max}}$ increases with $a/M$ (or $g/M$) up to a critical value of $a/M$ (or $g/M$), beyond which the corresponding spacetime does not have an event horizon for the chosen values of $q$ (or $p$).
This behaviour suggests that, for given $q$ (or $p$), the existence of a horizon imposes a natural upper bound on the extractable energy.
It is important to note that for finite values of $g$, the maximum efficiency $\delta E_{\text{max}}$ of the above class of rotating regular black holes is consistently lower than the maximum value for the $g=0$ case, which corresponds to the Kerr metric. Thus, at least for this class of regular black holes, one may
argue that regularisation of the metric leads to a lowering of the energy extraction 
efficiency.

\noindent The above result can also
be understood quite easily by an analytical argument. The essential point to note is that for the class of mass functions in (18), the radial
derivative of $m(r)$, i.e. $m'(r)$ is always greater than $0$ ($m'>0$).
Thus, if we evaluate the horizon condition $\Delta=r^2+a^2 - 2 m(r) r=0$
and its derivative $\Delta' = 2r -2m -2m'r=0$ at the extremal horizon $r=r_e$
we end up getting
\begin{eqnarray} 
\frac{a^2}{r_e^2} = 1- 2m'(r_e)
\end{eqnarray}
Since $m'>0$ for this class of mass functions ($g\neq 0$, regular)) and $m'=0$ corresponds to
the Kerr case ($g=0$, singular), the maximum extraction efficiency for the mass functions
$m(r)$ (assuming $E_{(0)}=1$ and the formula in (15),
will always be less than the maximum efficiency for Kerr.
\footnote{We thank the anonymous referee for suggesting this analytical argument.}

\section{A new rotating regular black hole and its energy extraction efficiency} \noindent In this section we examine the maximum energy extraction efficiency in a new class of rotating regular black holes, whose static version was first introduced in \cite{Kar1}. The static, spherically symmetric line element takes a form similar to Eq.(\ref{0.1}), with the following mass function,
\begin{eqnarray}\label{19}
    m(r)=\frac{1}{2}\frac{b_{0}^{2}r^{3}}{(r^{2}+g^{2})^{2}}
\end{eqnarray}
Here, $b_0$ is an arbitrary parameter of length dimension.
Note that the above mass function cannot be obtained from the generalised Neves-Saa metric for any choice of the parameters $p$ and $q$.
In the limit of vanishing regularisation parameter, it does not reduce to the Schwarzschild solution. Instead, it approaches the Reissner-Nordstr\"om type metric with zero mass and an `imaginary' charge ($Q^2 = -b_0^2< 0 $){\cite{Kar1}.
For small values of $r$, the metric becomes de Sitter spacetime, and asymptotically, it is a flat spacetime. For $0<g^{2}\leq 0.25\,b_0^2$, the geometry contains two horizons (inner and outer). When $g^{2}= 0.25\,b_0^2$, inner and outer horizons coalesce into a single horizon (extremal limit).
The matter required to support such a geometry violates the energy conditions (assuming GR) \cite{Kar1}. However, the supporting matter is modelled as a nonlinear electrodynamics sourced by a magnetic monopole, minimally coupled with GR. An alternative description of this spacetime in a braneworld scenario can also be found in Ref.\cite{Kar1}.

\noindent The rotating version of the geometry may be constructed following the same type-I prescription introduced by Bambi and Modesto (as discussed earlier).
Thus, the rotating counterpart may be expressed in terms of Eq.(\ref{0.3}) with the above mass function (Eq.\eqref{19}). When $g=0$ we obtain the
Kerr-Newman solution with $M=0$ and $Q^2=-b_0^2<0$ which is singular.
If $g\neq 0$ we obtain a regular geometry and its regular character can be established by showing that
all curvature invariants are finite in value throughout the domains of all coordinates. It is known that, seventeen Riemann curvature invariants can be constructed to characterise a general four-dimensional spacetime \cite{Zakhary}. However, a minimal independent set of  Zakhary-McIntosh (ZM) invariants for a Petrov type D spacetime with Segre type $[(1,1) (1 1)]$ includes the Ricci scalar, the Weyl invariants $I_1$, $I_2$, the Ricci invariant $I_6$, and the mixed invariants $I_9$, $I_{10}$ \cite{Zakhary}. 
For the sake of completeness, we provide the full expressions for five of them (excluding the Ricci scalar) below:
\begin{equation}
\begin{split}
    I_1&= C_{\alpha\beta\mu\nu}C^{\alpha\beta\mu\nu},\quad~ I_2= -C^{\alpha\beta\mu\nu}C^{*}_{\alpha\beta\mu\nu}, \quad~ I_6= R_{\mu\nu}R^{\mu\nu}, \quad~ 
    I_9= C_{\alpha \beta \mu}{}^{\nu} R^{\beta \mu}R_{\nu}{}^{\alpha},\\
    \text{and} & \quad~ I_{10}= -C^{*}{}_{\alpha \beta \mu}{}^{\nu} R^{\beta \mu}R_{\nu}{}^{\alpha},\\
    \end{split}
    \label{eqs:curvature_invariants}
\end{equation}
where $C_{\alpha\beta\mu\nu}$, $C^{*}_{\alpha\beta\mu\nu}$ are the Weyl tensor, the dual Weyl tensor respectively and are given as,
\begin{equation}
\begin{split}
    C_{\alpha\beta\mu\nu}&= R_{\alpha\beta\mu\nu} + \frac{R}{6} \left(g_{\alpha\mu}g_{\beta\nu}-  g_{\alpha\nu}g_{\beta\mu}\right) - \frac{1}{2} \left( g_{\alpha\mu}R_{\beta\nu}-  g_{\alpha\nu}R_{\beta\mu} - g_{\beta\mu}R_{\alpha\nu}+  g_{\beta\nu}R_{\alpha\mu}\right),\\
     C^{*}_{\alpha\beta\mu\nu}&= \frac{1}{2} \sqrt{-g} \epsilon_{\alpha \beta \rho \sigma} C^{\rho\sigma}{}_{\mu\nu}. 
\end{split}    
\end{equation}
All other Riemann curvature invariants can be expressed in terms of these six ZM invariants. Following \cite{Torres1, Kar3}, we have checked the finiteness of all the independent ZM invariants of this new rotating regular black hole spacetime. We found them to be finite everywhere.
It is expected that the required matter to support this new geometry will violate the standard energy conditions, as its static version does. We explicitly examine the energy conditions following Ref.~\cite {Kar3}. As the rotating metric has an off-diagonal term, it is difficult to directly use the standard energy condition inequalities for a purely diagonal metric. However, one may work in a chosen  Locally Non-rotating Frame (LNRF) with the following basis vectors, \cite{bardeen_LNRF}:
\begin{equation}
\textbf{e}^{(0)}= \left\vert- g_{tt} + \frac{g_{t\phi}^2}{g_{\phi\phi}} \right\vert^{1/2}\textbf{dt}, \quad~ \textbf{e}^{(1)}= \sqrt{g_{rr}}\,\textbf{dr}, \quad~ \textbf{e}^{(2)}=\sqrt{g_{\theta\theta}}\textbf{d}\,\theta, 
\quad~ \textbf{e}^{(3)}= \frac{g_{t\phi}}{\sqrt{g_{\phi\phi}}}\,\textbf{dt} + \sqrt{g_{\phi\phi}}\,\textbf{d}\phi.
\label{NLRF basis}
\end{equation}
In this basis, the required energy-momentum tensor is decomposed in the following way \cite{Poisson},
\begin{equation}
    T^{\alpha\beta}= \rho\, \hat{e}^{\alpha}{}_0\hat{e}^{\beta}{}_0 +p_1\, \hat{e}^{\alpha}{}_1\hat{e}^{\beta}{}_1 + p_2\, \hat{e}^{\alpha}{}_2\hat{e}^{\beta}{}_2 + p_3\,  \hat{e}^{\alpha}{}_3\hat{e}^{\beta}{}_3
\end{equation}
with $\rho$ being the energy density and $p_i$ ($i=\lbrace 1,2, 3 \rbrace$) the principal pressures given as 
\begin{equation}
    \rho = T^{\alpha\beta} \hat{e}_{\alpha}{}^0\hat{e}_{\beta}{}^0, \quad~ p_i= T^{\alpha\beta} \hat{e}_{\alpha}{}^i\hat{e}_{\beta}{}^i\,.
    \label{energy-pressure}
\end{equation}
$\hat{e}^{\alpha}{}_a$ are the inverse tetrad components which satisfies the orthogonality relation $\hat{e}_{\mu}{}^a\hat{e}^{\mu}{}_b=\delta^a_b$ (here Latin index $a$ represents the local inertial frame). Therefore, the energy conditions can be expressed in the LNRF frame in the following fashion,
\begin{equation}
    \begin{aligned}
       \text{Null energy condition:}& \hspace{0.2in} \rho+p_{i}\geq 0 \hspace{0.3in}(i=1,2,3)\\ 
       \text{Weak energy condition:}& \hspace{0.2in} \rho \geq 0, \hspace{0.2in} \rho+ p_{i}\geq 0 \hspace{0.3in}(i=1,2,3)
    \end{aligned}
\end{equation}
Here, we present the graphical analysis of the null and weak energy conditions of the new metric at $\theta=\pi/2$ for different parameter values.
\begin{figure}[h]
    \centering
    \includegraphics[width=0.326\textwidth]{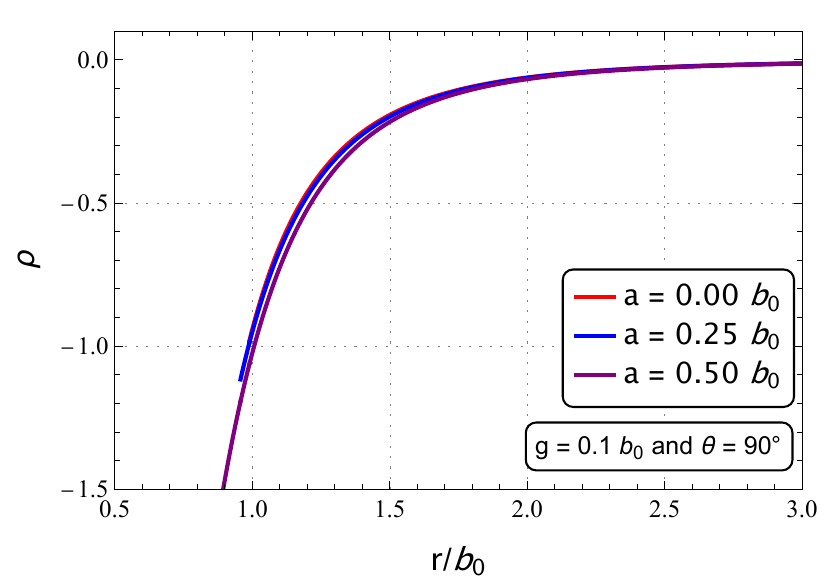}
         \includegraphics[width=0.326\textwidth]{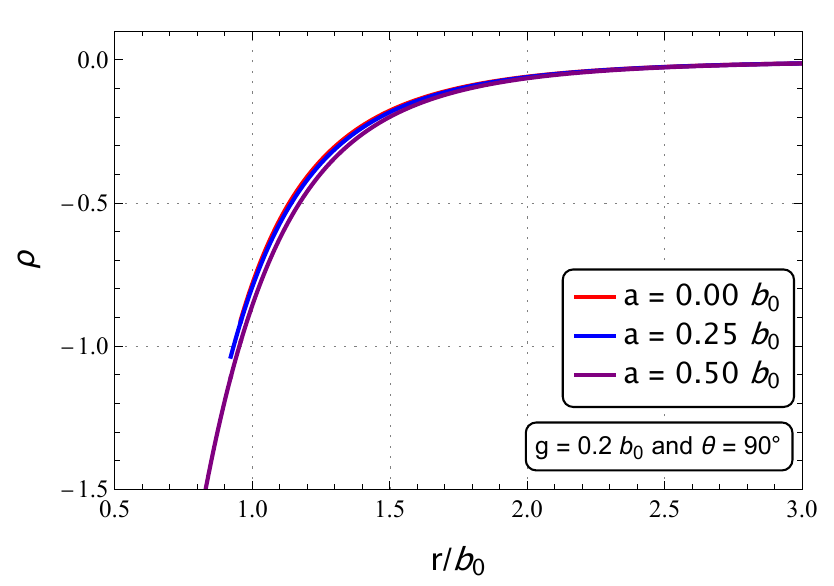}
         \includegraphics[width=0.326\textwidth]{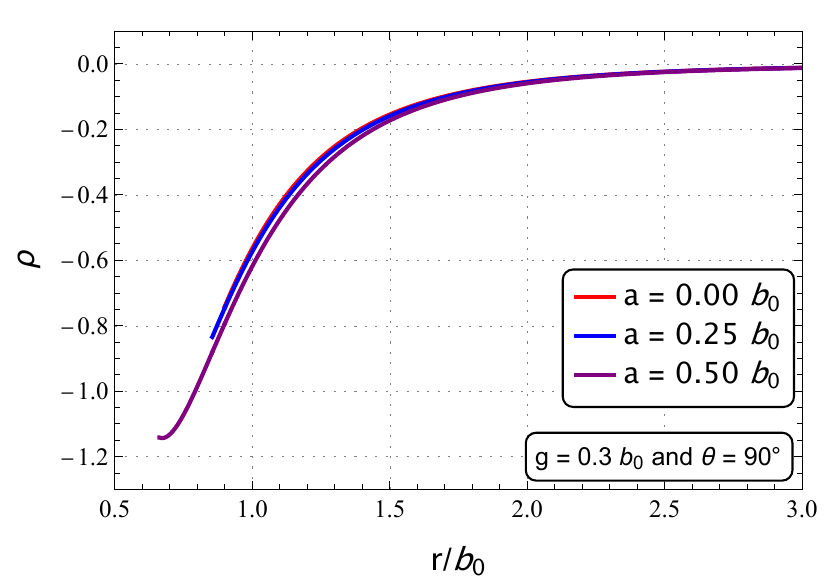}
    \caption{The variation of $\rho$ with $r$ for different values of $a$ and $g$ at $\theta=\pi/2$. Each curve started from its respective outer horizon radius. }
    \label{fig:WEC}
\end{figure}
In Figure \ref{fig:WEC}, the requirement of negative energy density for different values of $a$ and $g$ at $\theta=\pi/2$ indicates the violation of the weak energy condition. Thus, as with the Neves-Saa class of rotating solutions, the new metric also violates the weak energy condition.
To check the null energy condition, the different combinations of energy density and principal pressures are shown for various values of $a$ and $g$ at $\theta=\pi/2$, in Figure \ref{fig:NEC}.
\begin{figure}[h]
         \includegraphics[width=0.326\textwidth]{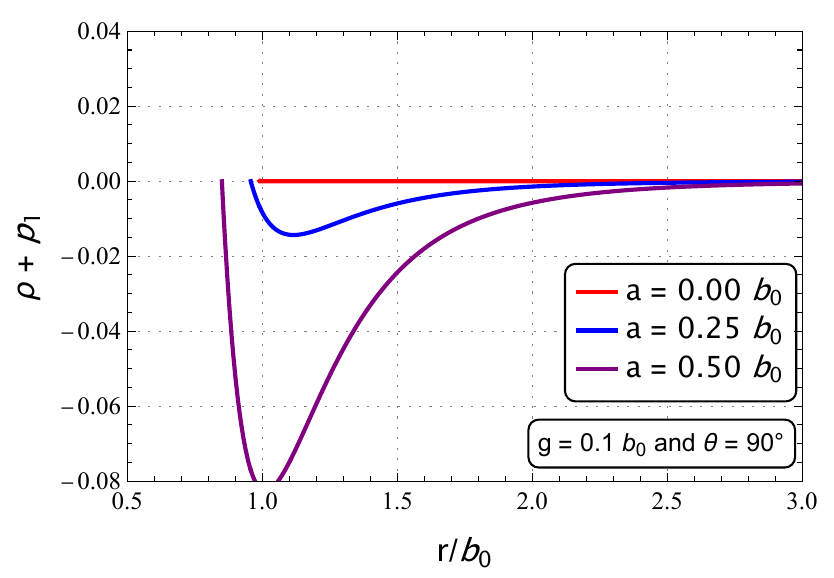}
         \includegraphics[width=0.326\textwidth]{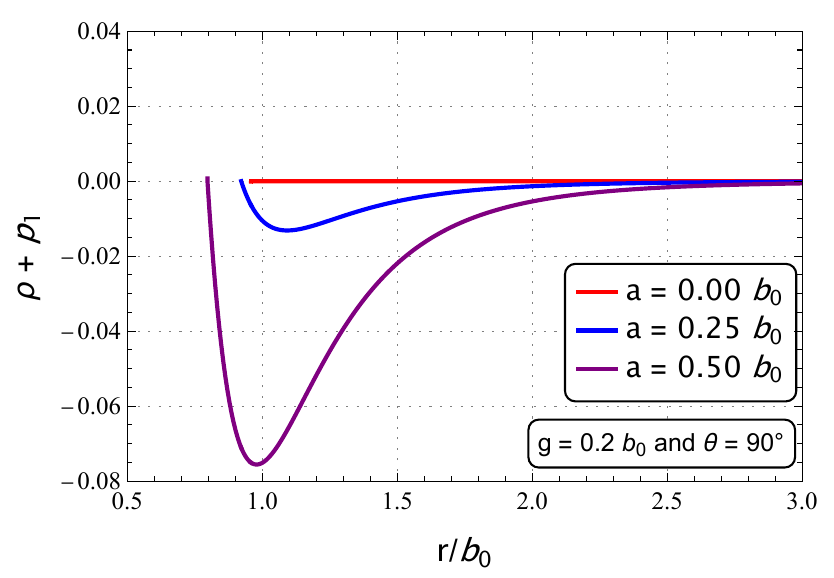}
         \includegraphics[width=0.326\textwidth]{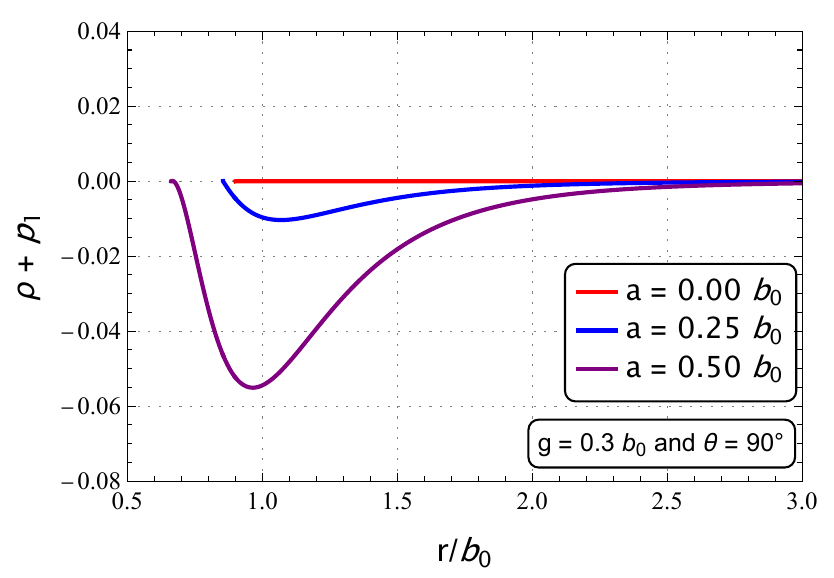}
         \includegraphics[width=0.326\textwidth]{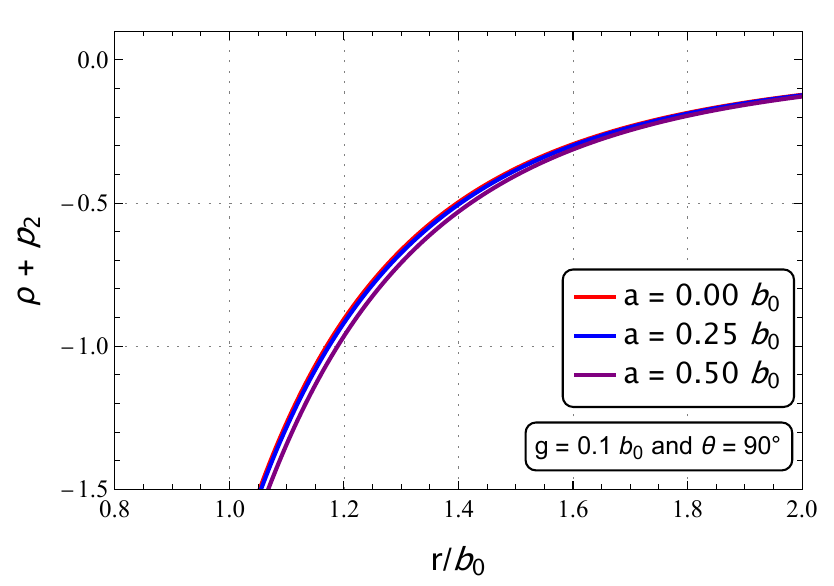}
         \includegraphics[width=0.326\textwidth]{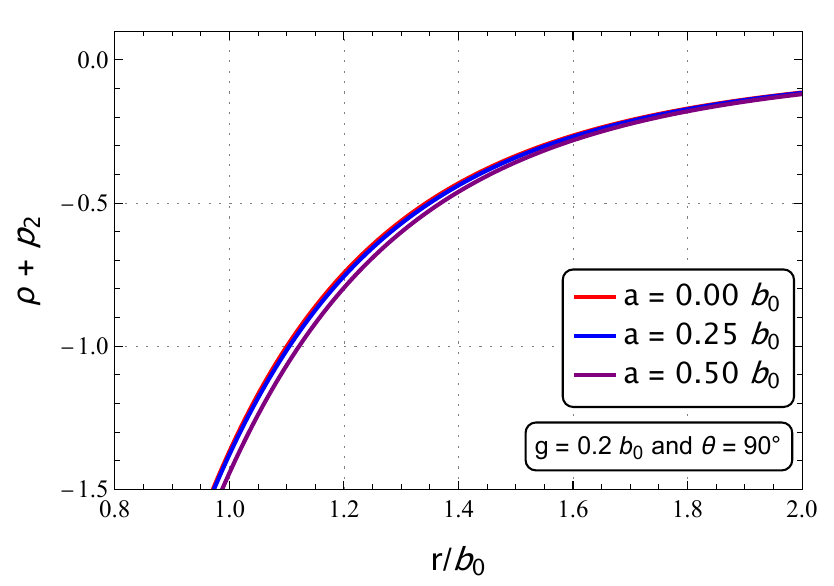}
         \includegraphics[width=0.326\textwidth]{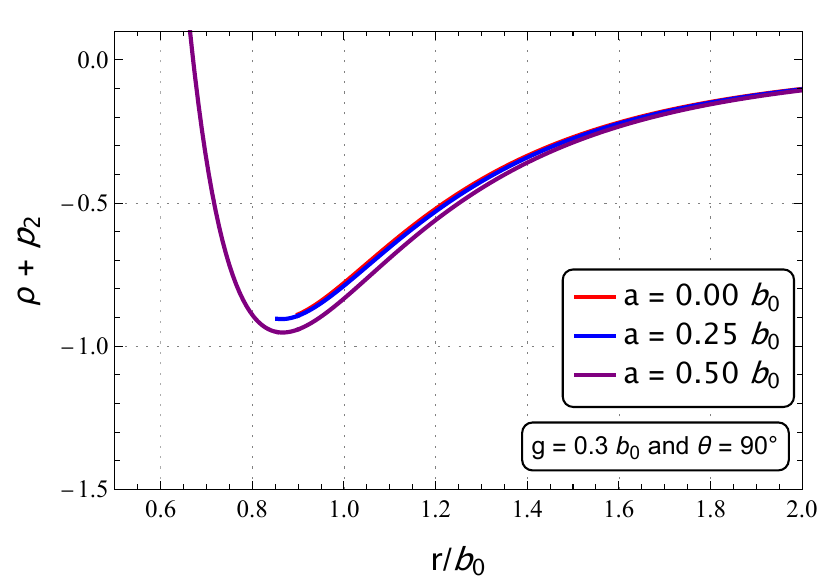}
         \includegraphics[width=0.326\textwidth]{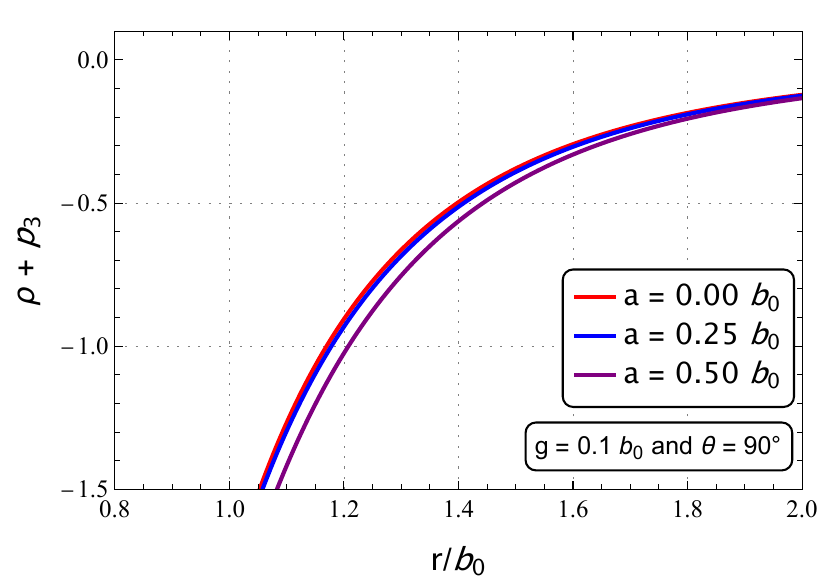}
         \includegraphics[width=0.326\textwidth]{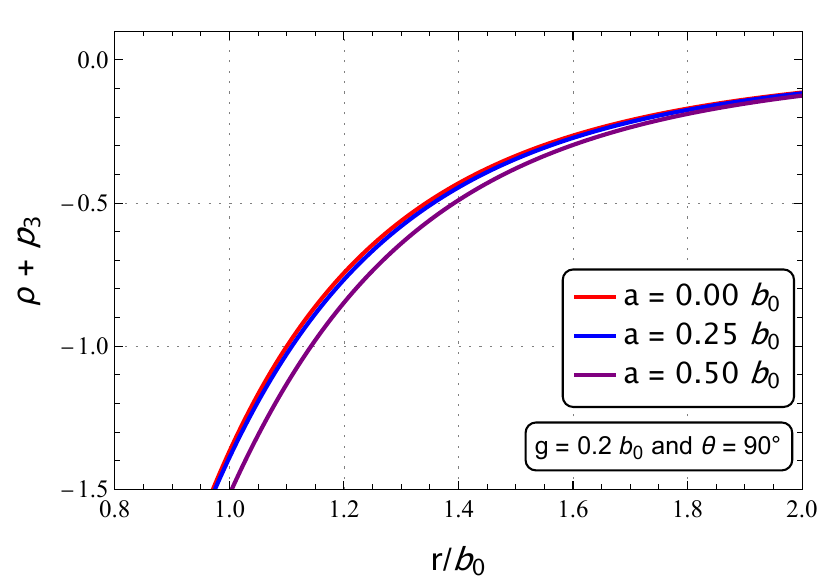}
         \includegraphics[width=0.326\textwidth]{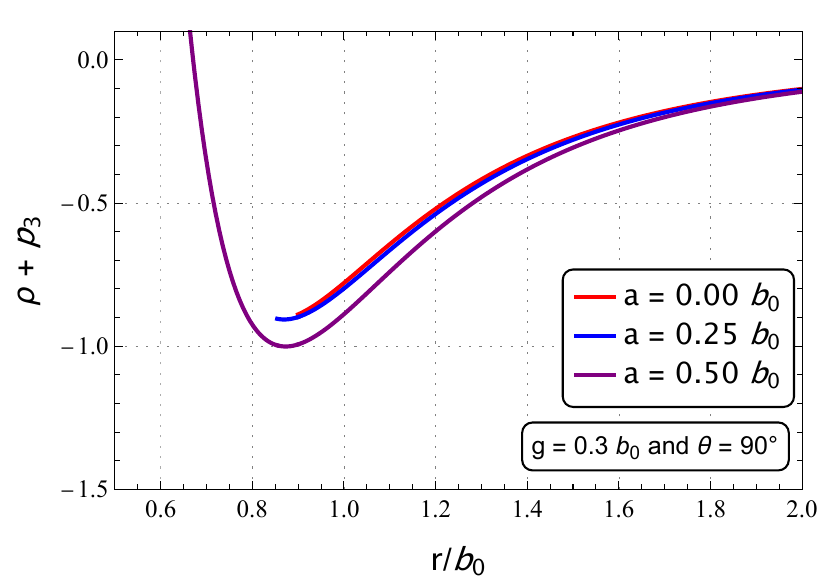}
        \caption{Plots of different null energy conditions with $r$ for different values of $a$ and $g$ at $\theta=\pi/2$. Each curve starts from the respective outer horizon radius given by the chosen $a$, $g$.}
        \label{fig:NEC}
\end{figure}ulae for likelihood-based tests of new physics
The three rows of Figure \ref{fig:NEC}, top, middle, and bottom, represent the variation of $(\rho+p_{1})$, $(\rho+p_{2})$, and $(\rho+p_{3})$, respectively.
We observe that the requirements of the null energy condition are not obeyed for different chosen sets of $(a,g)$. We plot each curve starting from the respective outer horizon radius.
The violation of the null energy condition further confirms that the other energy conditions will also be violated \cite{Visser}. Thus, the matter required to sustain the new rotating regular black hole violates all energy conditions, as long as we view the geometry as a solution in GR. 

\noindent However, it is possible that using the braneworld scenario, one may be able to circumvent such violations through the
purely geometric ${\cal E}_{\mu\nu}$ which depends on the bulk Weyl tensor,
and additional on-brane matter.  More specifically,  the effective Einstein equations in brane-world gravity \cite{shiromizu} are,
\begin{eqnarray}
G_{\mu\nu} = -\Lambda g_{\mu\nu}-{\cal E}_{\mu\nu} + \kappa T_{\mu\nu} + 6 \frac{\kappa}{\lambda}
S_{\mu\nu}
\end{eqnarray}
where $\Lambda$ is a four dimensional cosmological constant on the 3-brane, $T_{\mu\nu}$ is stress--energy of on-brane
matter and $S_{\mu\nu}$ is the so-called quadratic stress energy on the brane. The traceless ${\cal E}_{\mu\nu}$ is a purely
geometric quantity controlled by the extra dimension and related to the Weyl tensor of the
bulk five-dimensional spacetime. For our rotating geometry, one may include the energy-condition violating parts of matter in the traceless 
${\cal E}_{\mu\nu}$ and the rest can be adjusted in the additional terms involving on-brane matter.
This was done earlier for
the non-rotating case in \cite{Kar1}. The generalisation for the rotating version
seems plausible, but, is a fairly involved problem and we hope to address it separately, in future.

\noindent One may further note that,
though negative, the energy density and the pressures as well as their
combinations, are always bounded in value, as evident in the plots (Figure 9). 
Indirectly, this is an indication of the regular character of the spacetime
in which the avoidance of singularities is usually linked to a violation
of some of the assumptions (eg, energy conditions) in the Penrose-Hawking singularity theorems \cite{Wald}.

\noindent Similar to other rotating solutions discussed above, this new geometry also possesses two horizons 
with their locations depending on the parameter values.
\begin{figure}[!htbp]
\centering
\subfigure[\hspace{0.1cm}$g=0.4b_0$ and $a=0.1b_0$]{\includegraphics[width=0.326\textwidth]{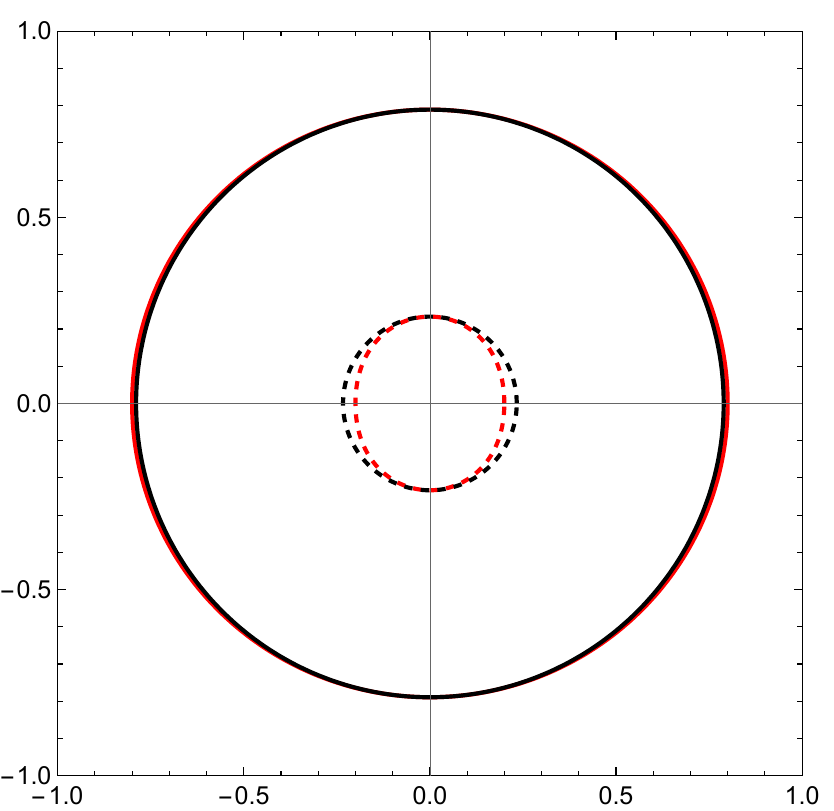}\label{subfig:K1}}
\subfigure[\hspace{0.1cm}$g=0.3b_0$ and $a=0.3b_0$]{\includegraphics[width=0.326\textwidth]{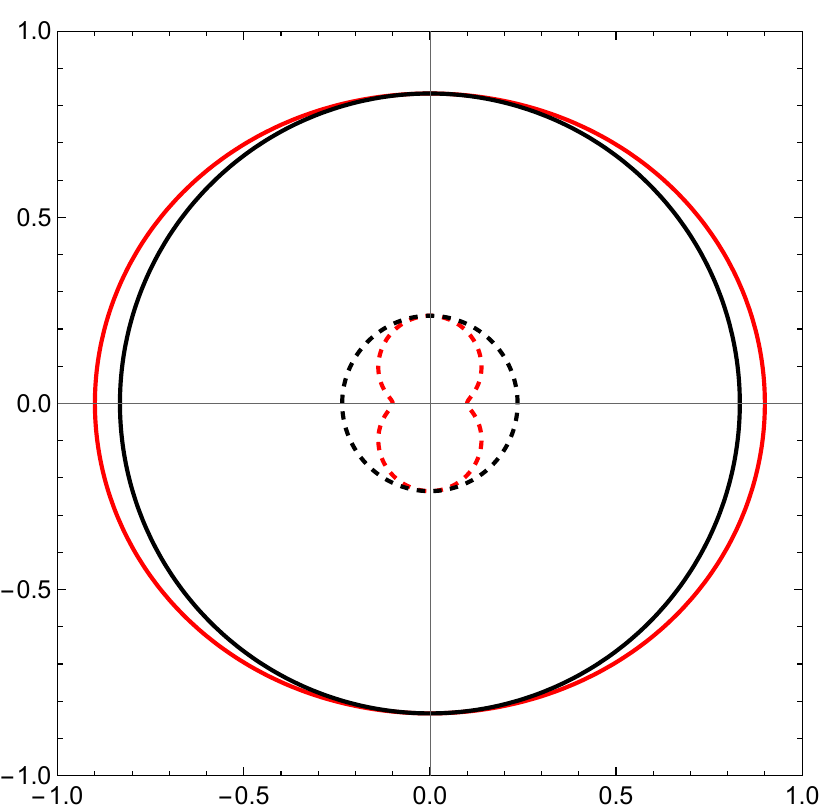}\label{subfig:K2}}
\subfigure[\hspace{0.1cm}$g=0.05b_0$ and $a=0.8b_0$]{\includegraphics[width=0.326\textwidth]{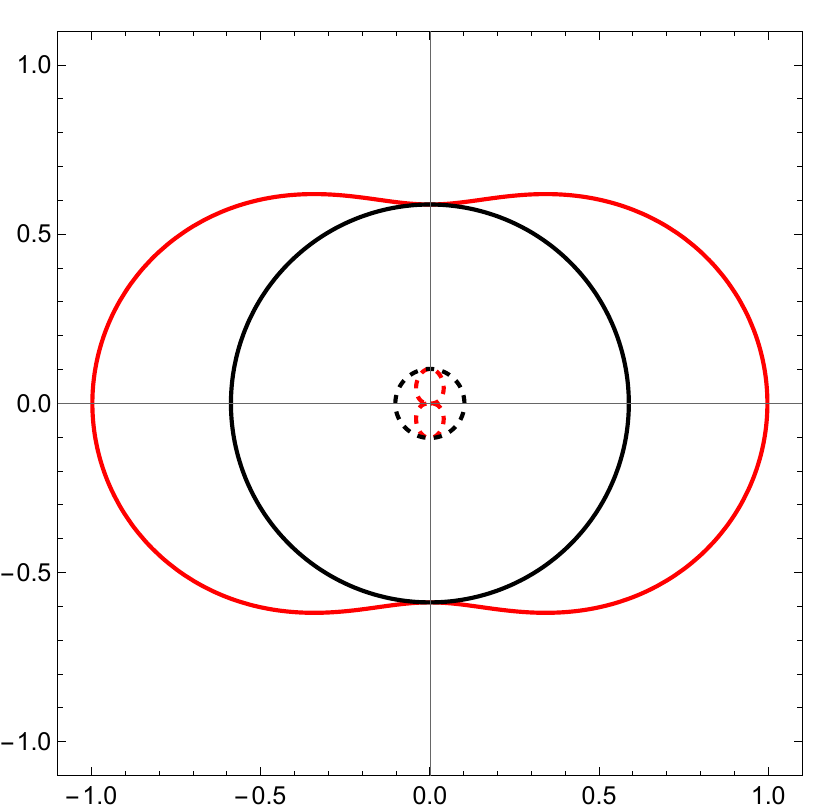}\label{subfig:K3}}
\caption{ Variation of ergoregion boundaries (red) and horizon radii (black) of the new rotating regular black hole for different pairs of $a$ and $g$}
\label{fig:KK}
\end{figure}
The shapes of the ergo-regions of the new geometry are shown in Figure \ref{fig:KK} for different values of $a$ and $g$. As expected, the width of the ergo-region grows with the spin parameter $a$. The presence of an ergo-region indicates that energy extraction is possible.
\begin{figure}[!htbp]
\centering
\subfigure[\hspace{0.1cm}Outer horizon]{\includegraphics[width=0.43\textwidth]{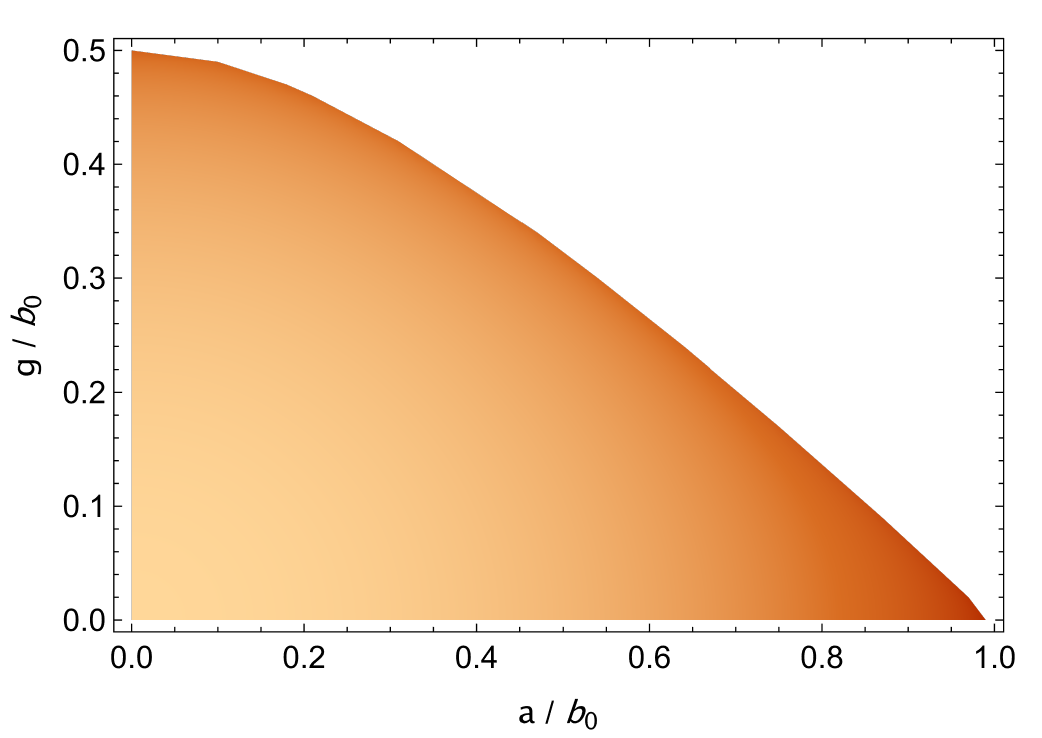}\label{subfig:K_outer}}
\subfigure[\hspace{0.1cm}Maximum efficiency]{\includegraphics[width=0.5\textwidth]{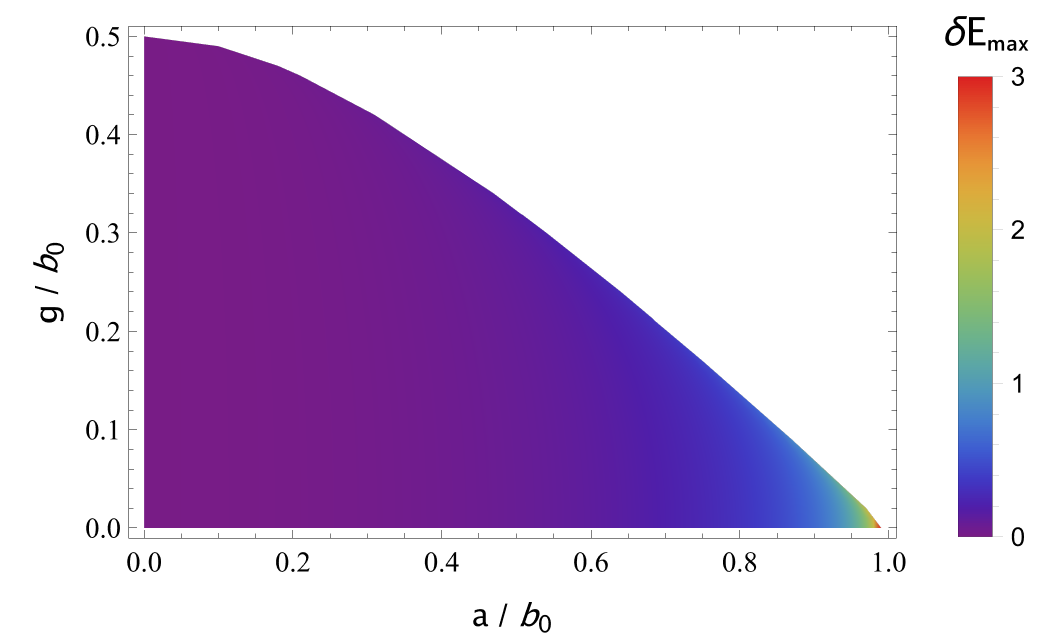}\label{subfig:K_extraction}}
\caption{The allowed values of $a/b_0$ and $g/b_0$ for which an outer horizon exists in the new rotating regular black hole (left). The right subfigure shows the variation of the maximum energy extraction efficiency with $a/b_0$ and $g/b_0$.}
\end{figure}
In Figure \ref{subfig:K_outer}, the allowed parameter space in terms of $a/b_0$ and $g/b_0$ is shown in colour, indicating the parameter values for which an outer horizon exists. The variation of maximum efficiency $(\delta E_{\text{max}})$ with $a/b_0$ and $g/b_0$ is presented in Figure \ref{subfig:K_extraction}. We find that for certain allowed values of the parameters, the maximum efficiency is greater than the Kerr case, though a comparison with the Kerr case may not be proper
since the geometry does not have a Kerr limit when $g=0$. A few selected values are listed in Table \ref{table1} as illustration. These values can also be obtained analytically
using (15) and by applying the method outlined earlier at the end of Section IV. The simplest case occurs when $g=0$ for which the
maximum efficiency can be found to be 3.04 by noting that $\frac{a^2}{r_h^2} = \frac{a^2}{b_0^2-a^2}$ and $\frac{a}{b_0} =0.99$, as noted in Table \ref{table1}. Earlier related work on the Penrose process in singular rotating 
braneworld black holes also indicate an enhanced efficiency as shown in
\cite{du}.

\begin{table}[h]
\centering
\begin{tabular}{ |c|c|c|} 
\hline
$a/b_0$ & $g/b_0$ & $\delta E_{\text{max}}$ \\
\hline
$0.99$ & $0.00$ & $3.04$ \\
\hline
$0.97$ & $0.02$ & $2.01$ \\
\hline
$0.96$ & $0.01$ & $1.31$ \\
\hline
$0.94$ & $0.03$ & $1.08$ \\
\hline
$0.91$ & $0.04$ & $0.77$ \\
\hline
$0.86$ & $0.09$ & $0.65$\\
\hline
\end{tabular}
\caption{A list of a few cases where $\delta E_{\text{max}}$ of the new rotating regular black hole exceeds that of the Kerr.}
\label{table1}
\end{table}
\noindent We observe that the energy extraction efficiency has its maximum when $g=0$ ($a$ is maximum) and decreases as $g$ increases. A similar behaviour was found for the Neves-Saa class of rotating solutions discussed
earlier.
This suggests that the parameter $g$ effectively suppresses the energy extraction efficiency. Although this trend is identical in both cases, the maximum efficiency depends on the singular seed geometry from which the regular
solution and its rotating regular counterpart are constructed, which is different between the two cases. 

 \noindent One may however ask why and how energy extraction 
becomes possible 
when the ADM mass and angular momentum are zero, as is the case for the rotating
geometry being discussed in this section. To understand why this happens, let us focus on the $g=0$ solution. 
We have already noted the horizon at $r_h=\sqrt{b_0^2-a^2}$ and the ergo-region
outer boundary (static limit surface) at $r_{s.l}^2 = b_0^2-a^2 \cos^2 \theta$.
The surface gravity turns out to be $\kappa = \frac{\sqrt{b_0^2 -a^2}}{b_0^2}$
and the horizon area is $A= 4\pi b_0^2$. These expressions seem to suggest the role of $b_0$ as somewhat similar to mass since for $a=0$ the surface gravity is $\frac{1}{b_0}$ (compare with Schwarzschild
where $\kappa=\frac{1}{4M}$). Using known general results nicely  reviewed in \cite{Poisson} 
and adapting to our case, it is possible to
show (laws of black hole mechanics) that,
\begin{eqnarray}
\delta M  - \Omega_H \delta J > 0
\end{eqnarray}
continues to hold. Here, $\delta M$ and $\delta J$ are the changes in the mass and
angular momentum of the black hole due to any external process and $\Omega_H= \frac{a}{b_0^2}$ (for $g=0$) is evaluated at the horizon.  We also require $\delta b_0>0$ which validates (28) through the increase in the
horizon area.
The above relation when applied with the energy ($\delta M$) and angular momentum ($\delta J$) of the infalling particle (after the Penrose process happens) 
changes the original stationary geometry to a new stationary geometry by 
altering its mass and angular momentum, even if $M$ and $J$ are zero to start with. This argument  can be generalised  to the cases
where $g\neq 0$ too, yielding the same generic result.
Hence, the basic logic leading to energy extraction
via the Penrose process is unaffected even if the ADM mass and angular momentum of the given original spacetime are zero. 

\noindent Alternatively, modifications of the metric function $f(r) = 1-\frac{2m(r)}{r}$
with additional terms (similar to the expressions in the $f(r)$ for Bardeen, Hayward and geometries in the Neves--Saa class), which have $\frac{1}{r}$ behaviour  at large $r$, will produce a non-zero ADM $M$ and $J$ for the
original stationary black hole and there too 
$\delta M - \Omega_H \delta J >0$ will continue to hold in the same way,
with the same conclusions w.r.t. the Penrose process.

\noindent It is also important to note the following fact about trajectories in this
regular rotating geometry. From the standard analysis of geodesics in the given spacetime, we obtain,
\begin{eqnarray}
    \left (\frac{dr}{d\tau}\right )^2 = \frac{b_0^2 r^2} {(r^2+g^2)^2} - \frac{{L^2_{(0)}}}{r^2} + \frac{a^2 (E_{(0)}^2-1)}{r^2}
    + \frac{b_0^2 (L_{(0)}-  E_{(0)}a)^2}{(r^2+g^2)^2} + (E_{(0)}^2-1)  
\end{eqnarray}

\begin{figure}[!htbp]
\centering
\subfigure[\hspace{0.1cm} $g=0$]{\includegraphics[width=0.4\textwidth]{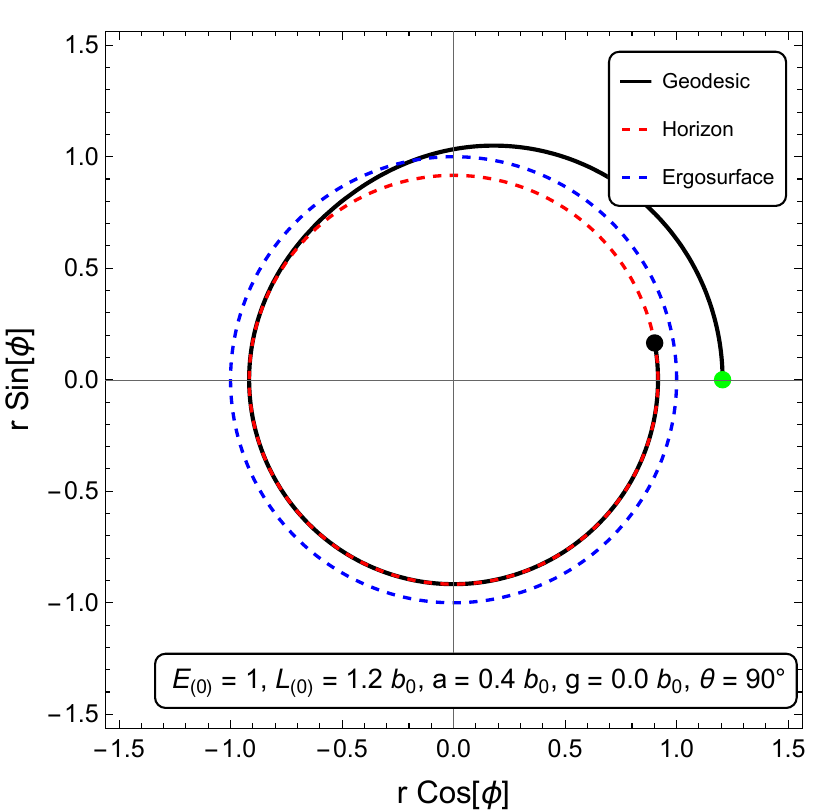}\label{subfig:Traj1}}
\subfigure[\hspace{0.1cm} $g\neq 0$]{\includegraphics[width=0.4\textwidth]{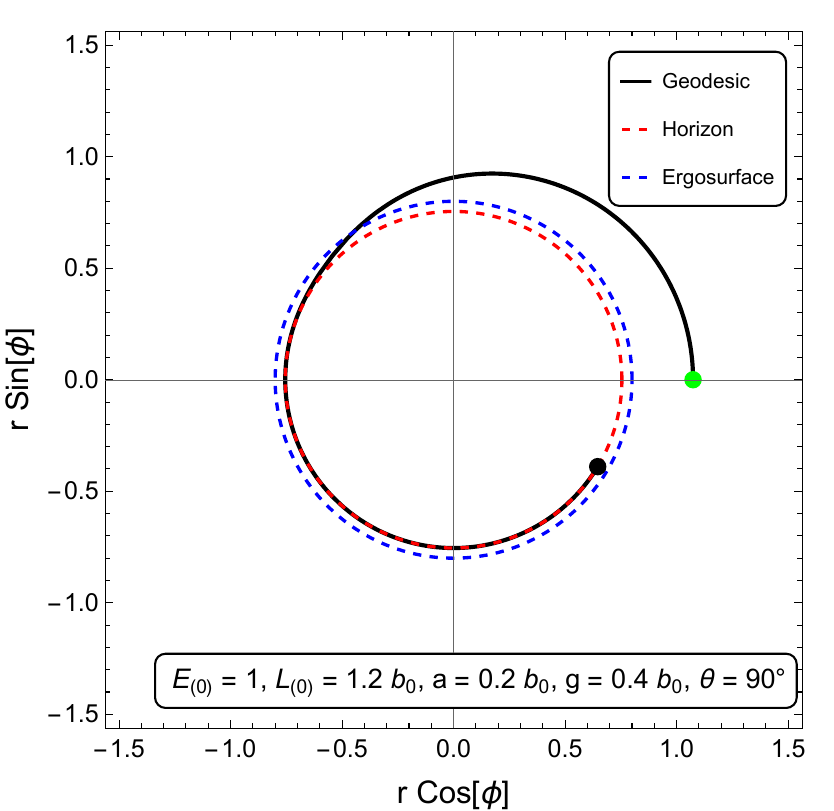}\label{subfig:Traj2}}
\caption{Trajectory plots for $g=0$ and $g\neq 0$ cases with the particle beginning at a finite distance outside the ergosurface and reaching the
vicinity of the horizon, where it  decays into two particles.}
\label{fig:Trajectories}
\end{figure}

\noindent For, $E_{(0)}=1$, the asymptotic ($r\rightarrow \infty$) form of the R. H. S. implies, from the
dominant $\frac{1}{r^2}$ term, that we must have $L_{(0)}^2< b_0^2$ if the 
particle starts out at radial infinity with zero velocity. 
If $E_{(0)}>1$, a similar analysis leads to $L_{(0)}^2< b_0^2 + a^2(E^2_{(0)}-1)$ with the particle now starting at infinity with a finite velocity.  
On the other hand, using $r\rightarrow r_h$ and the specific mass function in Eqn. (20), one would end up with
$L_{(0)} = \frac{r_h^2+a^2}{a} > b_0$. Thus, we seem to have
a contradiction for $E_{(0)}=1$. The way out for 
$E_{(0)}=1$ is to consider particle trajectories (timelike)
which are not coming from infinity--they could originate at a finite distance
(where the velocity is zero)
and reach the interior of the ergo-region where, near the horizon, the decay into two particles
may occur. In Figure 12, we plot the trajectories for $g=0$ and $g\neq 0$, assuming values such that $L_{(0)} = p b_0$ ($p>1$) and with $E_{(0)} =1$. The location from where the particle is dropped with
zero initial velocity is shown. One can see that such a particle
does reach the interior of the ergo-region where it begins to co-rotate
with the black hole and move towards the horizon. Thus, choosing $L_{(0)} > b_0$ it is still possible to build a scenario which can, theoretically, lead to energy
extraction via the Penrose process. For $E_{(0)} >1$, 
a window $b_0^2 <L_{(0)}^2 < b_0^2 + a^2 (E_{(0)}^2-1)$ exists for which the particle may start out at infinity but with a finite, non-zero velocity.  

\section{Discussion and conclusion}
\noindent In this paper, we have studied the maximum rotational energy extraction efficiency $(\delta E_{\text{max}})$ of rotating regular spacetimes via the Penrose process as a potential distinguishing feature between regular and singular black holes, entirely from a theoretical point of view. We first considered the rotating counterpart of the generalised Neves--Saa metric (parametrised by $p$ and $q$) as our working spacetime. We explicitly showed how the presence of a finite regularisation parameter $g$ leads to the reduction in the $\delta E_{\text{max}}$ for the rotating Bardeen and Fan-Wang metrics, which are special cases of the Neves--Saa metric. We also investigate $\delta E_{\text{max}}$ for other rotating solutions derived from the generalised Neves–Saa metric by varying the parameters $p$ and $q$ independently, while keeping other quantities fixed. We find that across the full spectrum of values of the parameters which define various special cases of this generalised Neves--Saa metric, the energy extraction efficiency always remains lower than that for the singular Kerr case. An analytical argument supporting this fact is
also derived.

\noindent Later, we construct, for the first time, the rotating version of a recently proposed static regular black hole metric and investigate the Penrose process and maximum energy extraction efficiency in this geometry. This new rotating, regular metric does not reduce to the Kerr metric under zero value of $g$ -- the regularisation parameter. Instead, it reduces to the Kerr--Newman type metric with vanishing mass parameter and an `imaginary' charge. 
Remarkably, we find that the $\delta E_{\text{max}}$ in such a metric does exceed the Kerr black hole value for some specific parameter choices. The maximum efficiency in this case occurs for
$g=0$ (singular geometry) and for $g>0$ (regular) the efficiencies are lower in value.

\noindent 
Our results suggest that the efficiency of the Penrose process, expectedly, is sensitive to the regular character of the chosen spacetime as well as the non-rotating seed geometry. Different classes of regular black holes can lead to notably different outcomes and may provide new insights into energy extraction beyond the Kerr case. A natural extension of the current work is to explore the repetitive Penrose process~\cite{Ruffini1, Ruffini2} in diverse rotating regular black hole spacetimes. We hope to communicate such work in the near future.
\\

\section*{Acknowledgement} 
\noindent AK is grateful for financial support through a PhD fellowship from the Indian Institute of Technology Kharagpur, India. SK thanks S. Shankaranarayanan
for discussions.

\section*{Data Availability Statement}
\noindent No datasets were generated or analysed during the current study.

\section*{Declarations}

\subsection*{Competing interests}
\noindent The authors declare that they have no competing interests.

\end{document}